\documentclass[a4paper,11pt]{article}
\pdfoutput=1 
\RequirePackage{snapshot}  
\usepackage{jheppub} 
\usepackage{bm,amsmath,amssymb,slashed,graphicx,%
            enumerate,alltt,xspace,multirow,xcolor,mathrsfs}
\usepackage{fancyvrb}
\usepackage{booktabs,tabularx}
\usepackage{graphicx}
\usepackage{subcaption}
\usepackage{xspace}
\usepackage{cancel}
\usepackage[utf8]{inputenc} 
\usepackage[compat=1.0.0]{tikz-feynman}
\usepackage{scalerel}
\usepackage{layouts}
\usepackage{adjustbox}
\usepackage{pdflscape}
\usepackage[normalem]{ulem} 
\usepackage{lineno}
\usepackage{wasysym}
\usepackage{relsize}
\usepackage{leftindex}
\usepackage{soul}
\usepackage{enumitem}
\usepackage{algorithm}
\usepackage[noEnd=true]{algpseudocodex}
\usepackage{placeins}
\renewcommand{\d}{\ensuremath{\mathrm{d}}}
\newcommand{\dPhi}{\ensuremath{\d\Phi}}
\newcommand{\eqRef}[1]{Eq.~\eqref{#1}\xspace}

\newcommand{\mc}[1]{\textsc{#1}\xspace}

\newcommand{\Alaric}{\mc{Alaric}}
\newcommand{\Ariadne}{\mc{Ariadne}}
\newcommand{\Dire}{\mc{Dire}}
\newcommand{\Apollo}{\mc{Apollo}}
\newcommand{\Deductor}{\mc{Deductor}}
\newcommand{\Herwig}{\mc{Herwig}}
\newcommand{\PanGlobal}{\mc{PanGlobal}}
\newcommand{\PanLocal}{\mc{PanLocal}}
\newcommand{\Pythia}{\mc{Pythia}}
\newcommand{\Sherpa}{\mc{Sherpa}}
\newcommand{\Vincia}{\mc{Vincia}}

\newcommand\betaps{\beta_\text{\textsc{ps}}}
\newcommand\betaobs{\beta_\text{obs}}
\newcommand{\mathd}{\mathrm{d}}

\newcommand{\ptilde}{{\widetilde p}}
\newcommand{\itilde}{{\tilde \imath}}
\newcommand{\jtilde}{{\tilde \jmath}}
\newcommand{\kperp}{k_{\perp}}

\newcommand{\calJ}{\mathcal{J}}

\usepackage[capitalise]{cleveref}

\makeatletter
\g@addto@macro\bfseries{\boldmath}
\makeatother

\definecolor{labelkey}{rgb}{0,0.5,0.0}
\definecolor{royalpurple}{rgb}{0.47, 0.32, 0.66}

\usepackage{listings}
\definecolor{darkred}{rgb}{0.6,0.0,0}
\definecolor{darkgreen}{rgb}{0,0.4,0}

\definecolor{grey}{rgb}{0.5,0.5,0.5}
\definecolor{rust}{rgb}{0.9,0.4,0.0}
\definecolor{lightblue}{rgb}{0.0,0.5,1.0}

\allowdisplaybreaks 

\definecolor{semiblue}{rgb}{0.3,0.3,0.8}
\newcommand{\logbook}[2]{}

\PassOptionsToPackage{hypertexnames=false}{jheppub}

\tikzstyle{block} = [rectangle, minimum width=1.0cm, minimum height=0.75cm, thin, draw=black]
\tikzstyle{blob} = [circle, minimum width=0.5cm, thin, draw=black]
\tikzset{blackarrow/.style={-stealth, semithick, draw=black}}
\tikzset{connection/.style={inner sep=0,outer sep=0}}

\newcolumntype{C}{>{\centering\arraybackslash}X}

\title{Timelike showers with jet recoils}

\newcommand{\Monashaff}{School of Physics and Astronomy, Monash University, Wellington Rd, Clayton VIC-3800, Australia}

\author[a]{Jack Helliwell,}%
\author[a]{Ludovic Scyboz,}%
\author[a]{Peter Skands}%

\affiliation[a]{\Monashaff}

\date{Received: date / Accepted: \today}

\abstract{
We propose a new way to impose four-momentum conservation on timelike parton-shower branchings, allowing for recoil to be imparted not only to individual partons but also to groups of partons, ``jets''.
In this work we present an explicit realisation of this idea for a dipole parton-shower, using angular ordering to decide which partons are grouped into jets in a way that does not require explicit jet clustering at each stage of the evolution.
We verify that the algorithm satisfies next-to-leading logarithmic (NLL) accuracy criteria, from numerical fixed-order tests as well as resummation tests across a range of observables. 
Our conclusion is that jet recoils provide a viable path for adapting existing dipole/antenna-type showers to achieve NLL accuracy.
}

\keywords{QCD, Parton Shower, Resummation, LEP
\\[4em]
\textit{For the purpose of Open Access, the authors have applied a CC BY
  public copyright licence to any Author Accepted Manuscript (AAM)
  version arising from this submission.}
}

\begin{document}

\maketitle
\section{Introduction}
\label{sec:introduction}

The question of recoils in parton showers has received much attention lately. This question arises because physical showers must conserve energy and momentum exactly:\footnote{See Ref.~\cite{Hoche:2017kst} for an interesting study exploring violations of this constraint.} any four-momentum imparted to an emitted parton must be balanced by momentum taken from elsewhere in the event; this balancing momentum is termed the \emph{recoil} from the branching process. Current methods to achieve this can be divided into two broad categories: \emph{local} and \emph{global}. 

In local recoil strategies, a single pair of partons takes the recoil. This is the minimum that is consistent with exact four-momentum conservation. Formally, the emission phase space is factorised as
\begin{equation}
\label{eq:psp-fact}
\dPhi_{n+1} ~=~ \dPhi_{n-2} \, \dPhi_{3} ~ = ~ \dPhi_{n} \,\frac{\dPhi_{3}}{\dPhi_2} \,,
\end{equation}
with two well-known examples being Catani-Seymour (CS)~\cite{Catani:1996vz,Phaf:2001gc} and antenna factorisation~\cite{Kosower:2003bh,Gehrmann-DeRidder:2009lyc}. Strategies based on these or similar principles have been employed successfully, e.g., in \Ariadne~\cite{Gustafson:1987rq,Lonnblad:1992tz}, \Pythia~\cite{Sjostrand:2004ef,Bierlich:2022pfr}, \Vincia~\cite{Fischer:2016vfv,Brooks:2020upa}, \Herwig's dipole showers~\cite{Platzer:2011bc,Bewick:2023tfi}, \Sherpa's CS shower module~\cite{Schumann:2007mg,Sherpa:2024mfk}, \Dire~\cite{Hoche:2015sya}, the \PanLocal showers~\cite{Dasgupta:2020fwr}, and others~\cite{Nagy:2006kb,Dinsdale:2007mf,Winter:2007ye}. 

In global strategies, some components of the recoil\footnote{At least the transverse component, but in some schemes also part of the longitudinal recoil.} are instead distributed among a wider subset of partons in the event (or simply even all partons in the event). Examples include \Herwig's angular-ordered showers~\cite{Marchesini:1983bm, Gieseke:2003rz,Bewick:2019rbu,Bewick:2023tfi}, the Manchester-Vienna dipole shower~\cite{Forshaw:2020wrq}, \Deductor~\cite{Nagy:2014mqa,Nagy:2020rmk}, and the \PanGlobal~\cite{Dasgupta:2020fwr,FerrarioRavasio:2023kyg}, \Alaric~\cite{Herren:2022jej,Assi:2023rbu,Hoche:2024dee}, and \Apollo~\cite{Preuss:2024vyu} showers. 

One virtue of local recoil schemes is simplicity: they can be straightforwardly nested within each other to build up higher-multiplicity phase spaces, 
and also exhibit manifest Lorentz invariance (i.e.~they are independent of any global choice of frame). The main disadvantage is that the transverse momentum imparted to one or both of the recoilers in these schemes can be  ``excessive''. In particular, for successive branchings that occur at commensurate transverse-momentum scales but at wide rapidity separations, local recoil schemes generally violate factorisation of independent emissions, and thus generate the wrong next-to-leading-logarithmic (NLL) terms~\cite{Dasgupta:2018nvj}. So far, the only known ways around this are to restrict the choice of shower evolution variable to a specific class that excludes  typical transverse-momentum-like ones~\cite{Dasgupta:2020fwr}, or to move to global recoils.\footnote{Angular ordering is a somewhat special case, which can achieve NLL for global observables at the expense of not filling parts of phase space that are important for non-global observables~\cite{Banfi:2006gy}. }

Global recoil schemes make it easier to ensure that no single parton receives excessive corrections from recoil effects. This enables such schemes to be made consistent with NLL~\cite{Dasgupta:2020fwr,Forshaw:2020wrq,Nagy:2020rmk,vanBeekveld:2022ukn,Herren:2022jej,Preuss:2024vyu} and higher~\cite{FerrarioRavasio:2023kyg,vanBeekveld:2024wws} accuracies for general evolution variables. Some disadvantages are that they tend to be more complicated:  the amount of operations needed to construct kinematics and update the event typically scales with the particle multiplicity squared $N^2$;
they may refer to a specific frame or reference vector and hence Lorentz invariance is less explicitly manifest; and the proof of the phase-space factorisation is more involved.

In this paper we propose a global approach that uses elements of local recoil schemes. It distributes recoil to dynamically-determined subsets of partons, or ``jets", in a way that 1) is based on the angular-ordering property of QCD, and 2) is compatible with NLL accuracy. The essential components of this scheme are a method to determine which particles --- the ``jet" --- will take the transverse recoil, and a (pre-existing) $2 \to 3$ kinematic map. The algorithm for determining the jets could also be used to define the global recoil vector in existing maps that leave this choice free, e.g.~the Manchester-Vienna shower~\cite{Forshaw:2020wrq}, \Alaric~\cite{Herren:2022jej}, and \Apollo~\cite{Preuss:2024vyu}.\footnote{We thank the referee for bringing this to our attention.} In this proof-of-concept work, we use a dipole map common to many parton-shower algorithms with dipole-local recoil, and take one of the momenta to be the jet instead of a single particle. We envisage that this strategy can be adapted to other $2\to3$ kinematic maps so that the showers that use  them also achieve NLL accuracy.
While we foresee that jet-recoil schemes may be formulated to embody the above advantages of local recoil schemes,
in the present work we focus on demonstrating the NLL accuracy of a jet-recoil scheme numerically, following the criteria of Refs.~\cite{Dasgupta:2020fwr, Dasgupta:2018nvj} at fixed order and all orders. We restrict our attention to final-state (timelike) showers with massless quarks;
generalisations to initial-state showers and to antenna kinematics will be explored in future studies. 



We note that, during the course of this work, we became aware of similar ideas being explored by Salam and Soyez in work that has not yet been published~\cite{SalamSoyez:unpublished}.

\section{Review of standard dipole/antenna showers}
\label{sec:standard-dipole}

A large set of parton showers are based on the dipole or antenna paradigm (i.e.~on the phase-space factorisation given in Eq.~(\ref{eq:psp-fact})). In this approach the event is represented as an ensemble
of coloured dipoles emitting incoherently (at leading colour, i.e.~in the large-$N_C$ limit). Consecutive emissions are generated in a Markov-chain Monte-Carlo
approach, ordered in some measure of hardness, the evolution variable, which is often taken to be a measure of the emission's transverse momentum, $k_t$, in the dipole centre-of-mass frame. The corresponding probability of going from a state $S_n$ consisting of $n$ partons to $S_{n+1}$, is given schematically by summing over colour-connected pairs $(\itilde,\jtilde)$:
\begin{equation}
\frac{\mathd\mathcal{P}(S_n)}{\mathd\ln k_t} = \sum_{\itilde,\jtilde} \int \mathd z \frac{\mathd \phi}{2\pi} \frac{\alpha_s}{\pi} P_{(\itilde \jtilde)}(k_t,z,\phi)\,,
\end{equation}
where $z$ is some longitudinal variable (e.g.~a momentum fraction), $\phi$ is an azimuthal angle, and $P_{(\itilde\jtilde)}$ represents an emission kernel --- a (partitioned) DGLAP or antenna function~\cite{Gribov:1972ri,Altarelli:1977zs,Dokshitzer:1977sg,Gustafson:1987rq,Catani:1996jh,Kosower:1997zr,Gehrmann-DeRidder:2005btv} ---  which in general may depend on all emission variables.

After each emission, the momenta of pre-existing particles must be modified so as to conserve the 4-momentum of the event as a whole. The exact way this is done is represented by the shower's specific kinematic map, $\mathbb{M} : S_n(\lbrace \ptilde\rbrace) \mapsto S_{n+1}(\lbrace p \rbrace)$. In \emph{local} dipole/antenna parton showers, 4-momentum is conserved within the emitting dipole,
\begin{equation}
\ptilde_l + \ptilde_m = p_l + p_m + p_k\,,
\end{equation}
with $\tilde p_{l}$ and $\tilde{p}_m$ the 4-momenta of the emitting dipole ends before the emission, and $p_l$, $p_m$, and $p_k$ the 4-momenta of, respectively, the two dipole ends and the emission, after the application of the map. That is, the phase-space factorisation expressed in \eqRef{eq:psp-fact} takes the explicit form,
\begin{equation}
\dPhi_{n+1} ~=~ \dPhi_{n-2} \, \dPhi_{lmk} ~ = ~ \dPhi_{n} \,\frac{\dPhi_{lmk}}{\dPhi_{\tilde{l}\tilde{m}}} ~.
\end{equation}

As we will review in section~\ref{sec:recoil-log-accuracy}, showers based on the principles of $k_t$-ordering and purely local recoil distribution in general do not produce the correct NLL resummation~\cite{Dasgupta:2018nvj,Dasgupta:2020fwr} (e.g.~for global event shapes). This means that in practice all NLL-accurate showers relinquish either the former or the latter property, or both.

\subsection{Sudakov parametrisation in dipole-local showers}

We first write the map $\mathbb{M}$ in the well-known Sudakov parametrisation to introduce some notation that will be useful when
we construct our new map.
For each iteration of the algorithm, standard (massless) dipole showers, generate an emission $k$ from a dipole stretching between two (pre-branching) colour-connected partons $\tilde{l},\tilde{m}$. The momentum of $k$ can be constructed as
\begin{equation}
\label{eq:dipole-map-pk}
p_k = a_k \tilde p_l + b_k \tilde p_m - \kperp \,,
\end{equation}
with a component $k_\perp$ being transverse to $\tilde p_l$ and $\tilde p_m $.
Here and below, we will choose to parametrise the coefficients $a_k$ and $b_k$ with the PanScales variables $k_t$, $\bar \eta$ (i.e.~$\betaps = 0$ in Ref.~\cite{Dasgupta:2020fwr}):
\begin{equation}
\label{eq:ak-bk}
a_k = \sqrt{\frac{s_{\jtilde}}{s_{\itilde} s_{\itilde \jtilde}}} k_t e^{\bar \eta}\,,\quad b_k = \sqrt{\frac{s_{\itilde}}{s_{\jtilde} s_{\itilde \jtilde}}} k_t e^{-\bar \eta}\,,
\end{equation}
where the Lorentz invariants are defined as $s_{ab}:=2 p_a \cdot p_b$ and $s_a := 2p_a \cdot Q$, with $Q$ the event's total 4-momentum.
The transverse component $\kperp$ is defined through two light-like 4-momenta $n_{\perp,1}$, $n_{\perp,2}$ such that $n_{\perp,1} \cdot n_{\perp,2} = n_{\perp,i}\cdot \tilde p_l = n_{\perp,i}\cdot \tilde p_m = 0$, and $n_{\perp,1}^2 = n_{\perp,2}^2 = -1$:
\begin{equation}
\label{eq:kperp}
    \kperp = k_t \left[ n_{\perp,1} \cos \phi + n_{\perp,2} \sin\phi \right]\,.
\end{equation} 

Showers with dipole-local recoil schemes directly ensure total 4-momentum conservation among $\lbrace l,m,k \rbrace$ by assigning transverse recoil exclusively to the emitter, say $l$, while the spectator, $m$, takes only a longitudinal component. That is, the (post-branching) emitter and spectator are given by:
\begin{align}
p_l &= a_l \tilde p_l + b_l \tilde p_m + \kperp\,,\label{eq:dipole-map-pi}  \\ 
p_m &= b_m \tilde p_m \,. \label{eq:dipole-map-pj} 
\end{align}

The coefficients $a_l, b_l, b_m$ are determined by on-shell conditions and conservation of the total 4-momentum. The emitter is typically taken to be the end of the dipole which is ``closer'' to the emission with respect to \emph{some} measure (e.g.~rapidity), sometimes with a smooth transition between the two dipole ends.
%

\subsection{Recoil distribution and logarithmic accuracy}
\label{sec:recoil-log-accuracy}

At all orders, we follow the usual logarithmic counting~\cite{Banfi:2001bz,Banfi:2004yd}: for observables $O$ that exponentiate, like global event shapes, the all-order series can be re-arranged~\cite{Catani:1992ua} to express the probability that the observable takes an exponentially small value,
\begin{equation}\label{eq:sigmaNLL}
\Sigma(\alpha_s, L) = P(O < e^L) = \exp \left[ L g_1(\alpha_s L) + g_2(\alpha_s L) + \mathcal{O}(\alpha_s^n L^{n-1}) \right] + \mathcal{O}(e^L)\,,
\end{equation}
where $L$ is a large negative number ($|L| \gg 1$), the function $g_1$ encodes \emph{leading-logarithmic} terms (of the form $\alpha_s^n L^{n+1}$), the function $g_2$ the \emph{next-to-leading-logarithmic} terms ($\alpha_s^n L^n$), and so on.

As shown in Refs.~\cite{Dasgupta:2018nvj,Dasgupta:2020fwr}, transverse-momentum ordered showers with dipole-local maps of the form defined by Eqs.~(\ref{eq:dipole-map-pk})--(\ref{eq:dipole-map-pj}) do not generally reproduce the correct function $g_2$ for exponentiating observables; i.e., they deviate from all-order resummations starting from NLL terms. This was shown to be a consequence of the fact that such showers do not reproduce the correct infrared (IR) limit of tree-level matrix elements for emissions that are singly-logarithmically separated in phase space (that is, emissions which have a large hierarchy in either transverse momentum $\ln k_t$, or rapidity $\eta \simeq -\ln \theta$). This in turn is due to an improper attribution of recoil: specifically, in standard $k_t$-ordered dipole showers, emissions happening \emph{later} in the shower evolution have the potential to disturb earlier emissions by relative $\mathcal{O}(1)$ amounts across logarithmically-large regions of phase space.
In other words, even though an earlier emission may have been generated according to the correct (factorised) matrix-element limit, the final state, after further emissions, no longer corresponds to that same matrix element.
The requirement that showers reproduce IR limits of multi-particle matrix elements was promoted in Ref.~\cite{Dasgupta:2018nvj} to a criterion for NLL accuracy.

\begin{figure}
    \centering
    \includegraphics[width=0.7\linewidth]{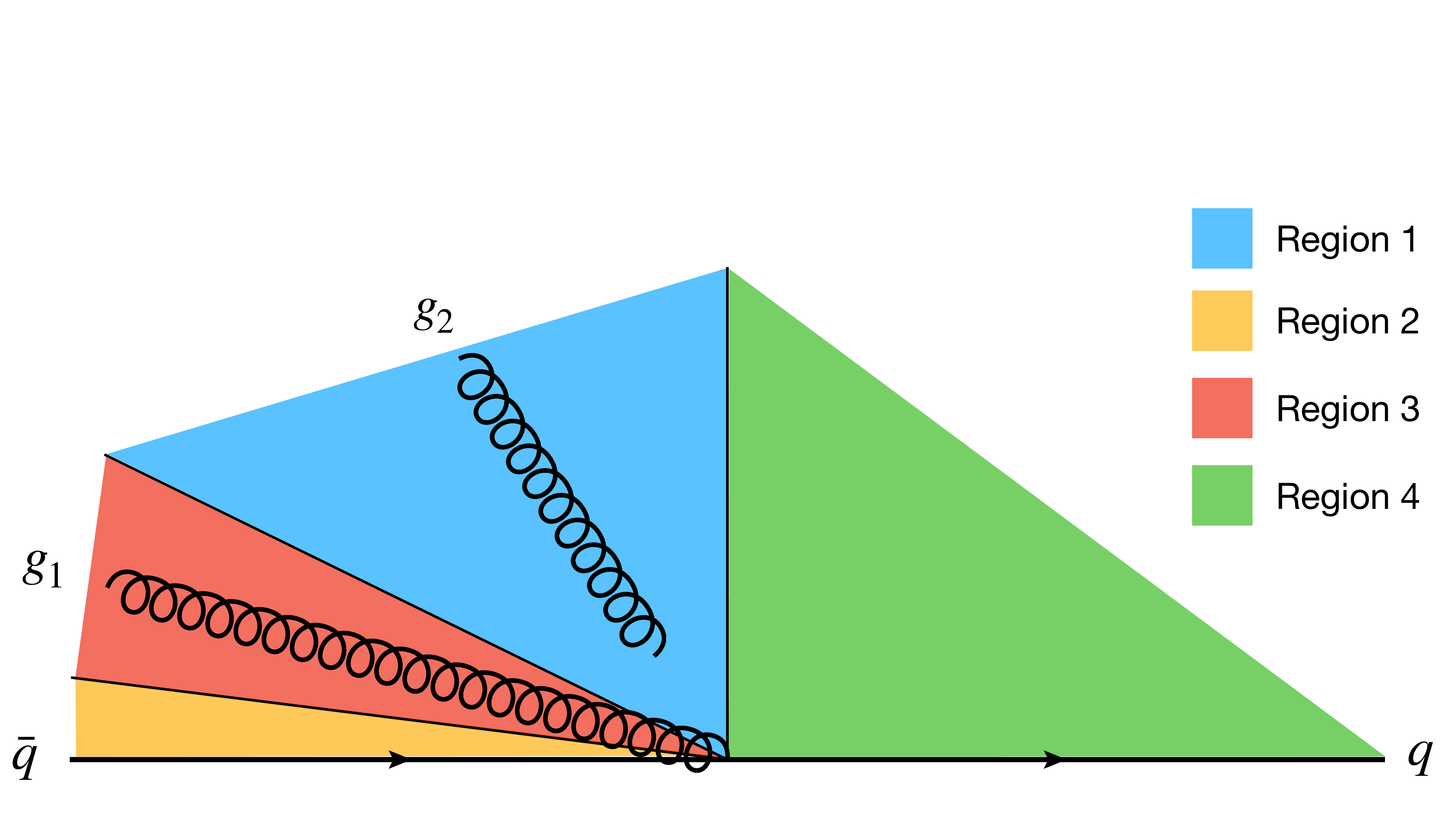}
    \caption{A three-parton system $\bar q q g_1$ emitting a gluon, $g_2$. Four distinct angular phase-space regions are highlighted, with the emission being shown in region 1.}
    \label{fig:2-gluon-configs}
\end{figure}

Consider an emission from a 3-parton ($q \bar q g_1$) system, where the first gluon is collinear to the anti-quark and soft ($\ln k_{t,1}/Q \ll 0$), as shown in Fig.~\ref{fig:2-gluon-configs}.
In a dipole shower, the second gluon $g_2$ can be emitted either from the $\bar q g_1$ or the $g_1 q$ dipole. Let us imagine that the second emission has a commensurate value of the evolution variable, $k_{t,2} \lesssim k_{t,1}$ and is emitted from the $g_1 q$ dipole. From factorisation we expect that for strongly-ordered angles, the two emissions are independent,
\begin{equation}
    |\mathcal{M}_{\rm PS}|^2 \simeq  \frac{C_F^2}{2} \frac{\alpha_s(k_{t,1}^2)\,\alpha_s(k_{t,2}^2)}{\pi^2} \frac{\mathrm{d}k_{t,1}^2\mathrm{d}k_{t,2}^2}{k_{t,1}^2k_{t,2}^2} \mathrm{d}\eta_1 \mathrm{d}\eta_2\,, \quad |\eta_1 - \eta_2| \gg 1\,,
\end{equation}
i.e.,~the second emission should not impact the kinematics of the first. In $k_t$-ordered showers, however, dipole-local maps give the transverse recoil to the first gluon $g_1$ over an erroneously large region of phase space, where the two gluons are emitted at very different angles.\footnote{This issue at $\mathcal{O}(\alpha_s^2)$ was recognised early on and a fix that gives the recoil only to the quark was implemented in \Ariadne~\cite{Andersson:1991he,Lonnblad:1992tz}. Nevertheless the issue is simply postponed to one order higher~\cite{Dasgupta:2018nvj}.} This results in the shower failing to reproduce the correct IR tree level matrix element for two emissions at commensurate $k_t$, but far apart in rapidity. The incorrect matrix element is produced over a logarithmically-enhanced region of width $\propto \eta_1$ (for a fixed $k_{t,1} \sim k_{t,2}$), and thus will contribute incorrectly to the single-logarithmic coefficient at that order. 

Whether or not subsequent emissions spuriously modify the kinematics of prior emissions can be tested at fixed order~\cite{Dasgupta:2018nvj,vanBeekveld:2022zhl,vanBeekveld:2023chs} by examining distributions on a Lund diagram~\cite{Andersson:1988gp,Dreyer:2018nbf}, which represents the phase space available to the next emission in terms of $\ln k_t$ and $\eta$.
Requiring that all emissions are separated by a distance $L$ on that plane, the discrepancy between the shower and the IR limit of tree-level matrix elements should then vanish as $e^{-|L|}$. 
For this to be true, recoil effects must also vanish exponentially (or faster) with the distance $L$.

We begin with a base event, consisting of a Born $q \bar{q}$ pair along with one or more emissions.
This event is then first clustered to two hemispheres using the Cambridge algorithm~\cite{Dokshitzer:1997in} (which is implemented in FastJet~\cite{Cacciari:2011ma}). Each hemisphere corresponds to one half of the primary Lund plane. If all emissions are soft and/or collinear then the one hemisphere will contain the quark, and the other the anti-quark. 
The Lund plane can then be constructed recursively as follows~\cite{Dreyer:2018nbf}:
undo the last step in the clustering sequence and determine the transverse momentum and rapidity of the softer branch with respect to the harder branch. These coordinates are recorded on the relevant half of the primary Lund plane.
Repeat this procedure, following the harder branch, until it has been fully de-clustered. The secondary Lund leaves (phase-space regions associated with each emission) can then be constructed by repeating the same procedure but following, in turn, each softer branch.

\begin{figure}
    \centering
    \includegraphics[width=0.6\linewidth,page=7,trim={0 0 0 1cm}, clip]{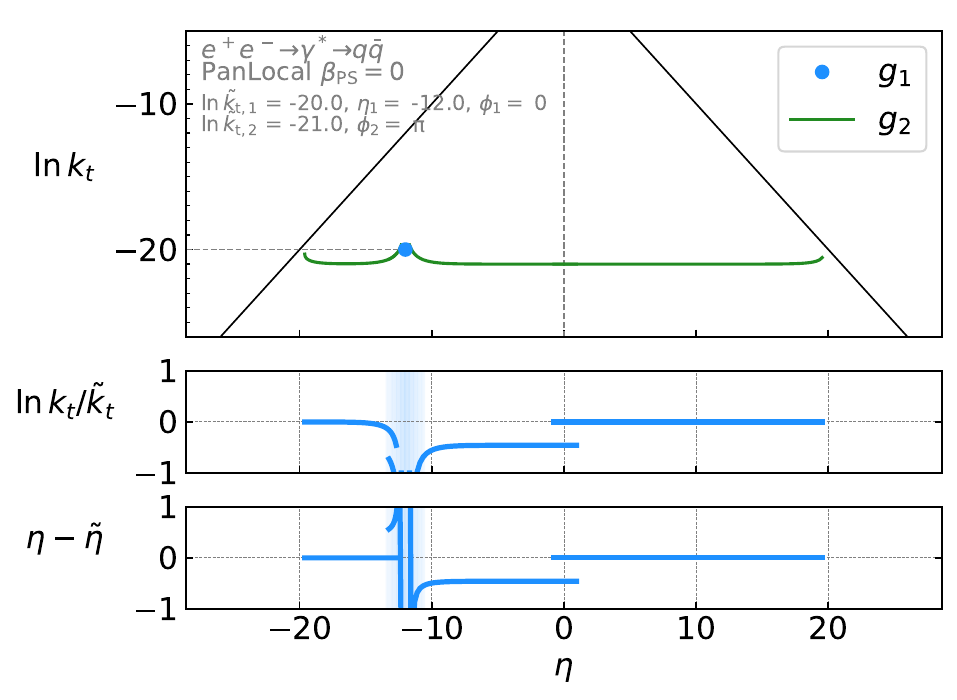}
    \caption{Illustration, as in Refs.~\cite{Dasgupta:2018nvj,vanBeekveld:2022zhl}, of incorrect recoil assignment for PanLocal ($\betaps = 0$), which leads to the wrong NLL terms (the PanLocal shower is NLL-accurate for $\betaps > 0$). The top panel shows the Lund plane  coordinates of a first emission $g_1$ and the contour along which a second emission $g_2$ is produced, in green. The lower two panels show the change in the first emission's transverse momentum ($\ln k_t / \tilde k_t$) and rapidity ($\eta-\tilde \eta$) due to the emission of $g_2$. The large recoil given in region $\eta_1 < \eta < 0$ is problematic at NLL, whereas the shaded region around $\eta \simeq \eta_1$ contributes only at NNLL.}
    \label{fig:wrong-NLL-fo}
\end{figure}

For the purposes of these fixed-order tests, having constructed the Lund diagram for the base event, an additional emission is then generated at a specific $(k_t, \, \phi)$ and scanned across the available range of the third kinematic variable, i.e., rapidity $\eta$. Any change in the transverse momentum or rapidity of prior emissions is then recorded as a function of the phase-space coordinates of the new emission. An example of the result of such a test is shown, for the configuration described above, in Fig.~\ref{fig:wrong-NLL-fo} using the PanLocal shower (similarly to Refs.~\cite{Dasgupta:2018nvj,vanBeekveld:2022zhl,vanBeekveld:2023chs}). This shower achieves NLL accuracy only for $\betaps >0$, which excludes transverse-momentum ordering ($\betaps = 0$). In Fig.~\ref{fig:wrong-NLL-fo}, we purposefully set $\betaps=0$ to illustrate the fixed-order behaviour that gives rise to the wrong NLL terms. In the two lower ratio plots, one can see that both the transverse momentum and rapidity of $g_1$ are modified by the production of $g_2$ when $\eta_1 < \eta_2 <0$ (i.e.,~a logarithmically-enhanced region), which corresponds to region 1 in Fig.~\ref{fig:2-gluon-configs}. In contrast, the large change in the momentum of $g_1$ when the two gluons are at commensurate transverse momentum \emph{and} rapidity (the blue shaded region in the ratio plots) only becomes important starting from NNLL onwards~\cite{Dasgupta:2020fwr,FerrarioRavasio:2023kyg,vanBeekveld:2024wws}.
%

\section{A jet-based recoil algorithm}
\label{sec:shower-desc}

In this section we will formulate a new prescription for the attribution of recoil, that we dub ``jet recoil", which assigns transverse recoil in a semi-global manner such that multiple-emission matrix elements are reproduced to the extent required for the shower to achieve NLL accuracy. Our guiding principle will be coherence~\cite{Mueller:1981ex,Mueller:1983js,Ermolaev:1981cm,Bassetto:1982ma,Dokshitzer:1982xr,Dokshitzer:1982ia}, which we will see lifts the issue of excessive recoil that leads to the wrong NLL terms, as discussed in the previous section.\footnote{Ref.~\cite{Hamilton:2020rcu} already noted the possibility that one could design a shower map that follows the same pattern as the colour-factor assignment. The Manchester-Vienna shower~\cite{Forshaw:2020wrq} also makes mention of angular ordering as a guiding principle in the construction of the shower.}

To illustrate the principle, we first work through a few examples for gluon emissions from a $\bar q q g_1$ event in section~\ref{sec:example-as2}. The general procedure for determining the recoiling jet from a set of intervals, stored alongside the list of emitting dipoles, is then given in section~\ref{sec:jet-finding}, as well as that for updating the list of intervals after each emission in section~\ref{sec:interval-update}. We will then explain how we handle special cases (in particular $g\to q \bar q$ splittings) in section~\ref{sec:edge-cases}.

\subsection{An example: emissions up to $\mathcal{O}(\alpha_s^2)$}
\label{sec:example-as2}

In our shower, we will always follow the same construction as in the usual leading-colour dipole showers to generate an emission: as a first step, we generate an emission $k$ from a dipole stretching between two colour-connected ends, $(\tilde l,\tilde m)$, according to Eq.~(\ref{eq:dipole-map-pk}). We do not, however, construct the post-branching momenta $(l,m)$ as in Eqs.~(\ref{eq:dipole-map-pi})--(\ref{eq:dipole-map-pj}), but rather distribute the recoil, in a second step, to a jet $\calJ$ (which absorbs the transverse component $k_\perp$) and a spectator (which only receives longitudinal recoil) following the procedure below. For the purpose of kinematics, the jet is akin to the ``emitter" in a standard dipole-local map.

In the example of the previous section (region 1 of Fig.~\ref{fig:2-gluon-configs}), the opening angle $\theta_{1\bar{q}}$ between $g_1$ and $\bar{q}$ is smaller than either of their opening angles to $g_2$, $\mathrm{min}(\theta_{12},\theta_{2\bar{q}})$; QCD coherence would then imply that $g_2$ should not be able to resolve $g_1$ and $\bar{q}$ as separate particles. Instead, $g_2$ should be emitted coherently from the sum of $\bar{q} + g_1$. Based on this notion we would prefer to assign the transverse recoil to a ``jet", $\calJ$, made up of $\lbrace \bar{q}, g_1 \rbrace$.
If the recoil is distributed amongst the jet constituents in such a way that the relative transverse momentum of the anti-quark and $g_1$ is preserved, then the logarithmic accuracy of the shower will not be adversely affected.
Starting from the aforementioned three-parton configuration, consider that the emission $g_2$ is generated from the $g_1 q$ dipole with $\eta_1 \ll \eta_2 < 0$ as in Fig.~\ref{fig:wrong-NLL-fo}. According to the argument above we identify the recoiling jet as $\calJ = \{\bar q,g_1\}$ with momentum
\begin{equation}
    \tilde p_{\calJ}=\sum_{i\in \calJ} \tilde p_i\ .
\end{equation}
We will provide a generic algorithm for determining $\calJ$ in section~\ref{sec:jet-finding}. 
We additionally choose the spectator to be the quark, i.e.,~the end of the emitting colour dipole farthest in rapidity from the emission.\footnote{The spectator $s$ is therefore either $l$ or $m$, depending on which dipole end the emission is most collinear to. We believe that there is considerable freedom in how one chooses the spectator, or even promote it to be a jet itself.} The post-branching jet and spectator momenta can then be constructed~\cite{Phaf:2001gc} and are given by
\begin{align}
    p_\calJ &= a_{\calJ}\, \ptilde_{\calJ} + b_{\calJ}\, \ptilde_s +k_{\perp}\,,\\
    p_s &= b_s\, \ptilde_s\, ,
\end{align}
where the coefficients of the map are given by
\begin{align}
a_{\calJ} =& 1-\frac{s_{\tilde{s}k}}{s_{\tilde{\calJ}\tilde{s}} }\,,\\
b_{\calJ} =& \frac{s_{\tilde{s}k}}{s_{\tilde{\calJ}\tilde{s}}}\left(2\xi^2+\frac{s_{\tilde{\calJ}k}}{s_{\tilde{\calJ}\tilde{s}}-s_{\tilde{s}k}}\right)\,,\\
b_s  =& 1 - \left(\frac{s_{\tilde{\calJ}k} }{ s_{\tilde{\calJ}\tilde{s}} - s_{\tilde{s}k} } \right) \,,
\end{align}
with $\xi^2 = m_{\tilde{\calJ}}^2 / s_{\tilde{\calJ}\tilde{s}}$. Finally, we construct a Lorentz transformation $\Lambda$ such that $\Lambda \ptilde_\calJ = p_\calJ$, and apply it to all particles comprising $\calJ$:
\begin{equation}
    p_i = \Lambda\, \tilde p_i\,, \quad \forall\,i \in \calJ\,,
\end{equation}
thus preserving the invariants between pairs of particles within the jet.
The transformation applied to a 4-momentum $q$ can be broken into two successive transformations: a rotation,
\begin{equation}
\vec{q}\,' = \vec{q} + (1-\cos\theta) \left[ (\vec{q}\cdot \vec{n})\, \vec{n} -\vec{q} \right]
- \sin\theta (\vec{q}\times \vec{n})\,,
\end{equation}
where $\vec{n} = \frac{{\ptilde}_{\calJ} \times p_{\calJ} }{ \| \ptilde_{\calJ} \times p_{\calJ}\|}$ and $\theta$ is the opening angle between $\ptilde_{\calJ}$ and $p_\calJ$ in the event frame, followed by a boost to $q' = (q^0, \vec{q}\,')$ along the direction of $p_{\calJ}$ with velocity
\begin{equation}
\beta = \frac{E_{\tilde{\calJ}} \sqrt{E_{\tilde{\calJ}}^2 - m_{\calJ}^2 } - E_{\calJ} \sqrt{E_{\calJ}^2 - m_{\calJ}^2 }}{( E_{\tilde{\calJ}}^2 + E_{\calJ}^2 + m_{\calJ}^2)} \, .
\end{equation}
This prescription predominantly assigns the recoil to the hardest particles in the jet.

We now consider the situation where the second emission $g_2$ is much more collinear with the anti-quark than the first emission, $\eta_2 \ll \eta_1 \ll 0$, i.e.~where the gluon would be emitted in region 2 in Fig.~\ref{fig:2-gluon-configs}. This configuration would typically be associated with emission from the $\bar{q}g_1$ dipole. In this case the physical picture is that of the anti-quark splitting $\bar{q}\to \bar{q} g_2$ and so $\calJ$ should be composed only of the anti-quark, with $g_1$ acting as the spectator.

The final scenario to consider is where $g_2$ is more collinear to $g_1$ than to the anti-quark (region 3 of Fig.~\ref{fig:2-gluon-configs}), in which case we can view the process as $g_1\to g_1 g_2$, with the transverse recoil correspondingly given to $g_1$. In other words the jet is composed only of $g_1$. Again, we will choose the spectator to be the end of the emitting dipole that is farthest in rapidity from the emission, i.e, $q$ if the gluon is emitted from the $g_1 q$ colour dipole (and $\bar{q}$ if it is emitted from the $g_1\bar{q}$ one).

\begin{figure}[ht]
    \centering
    \includegraphics[width=0.6\textwidth,page=1,trim={0 0 0 1cm}, clip]{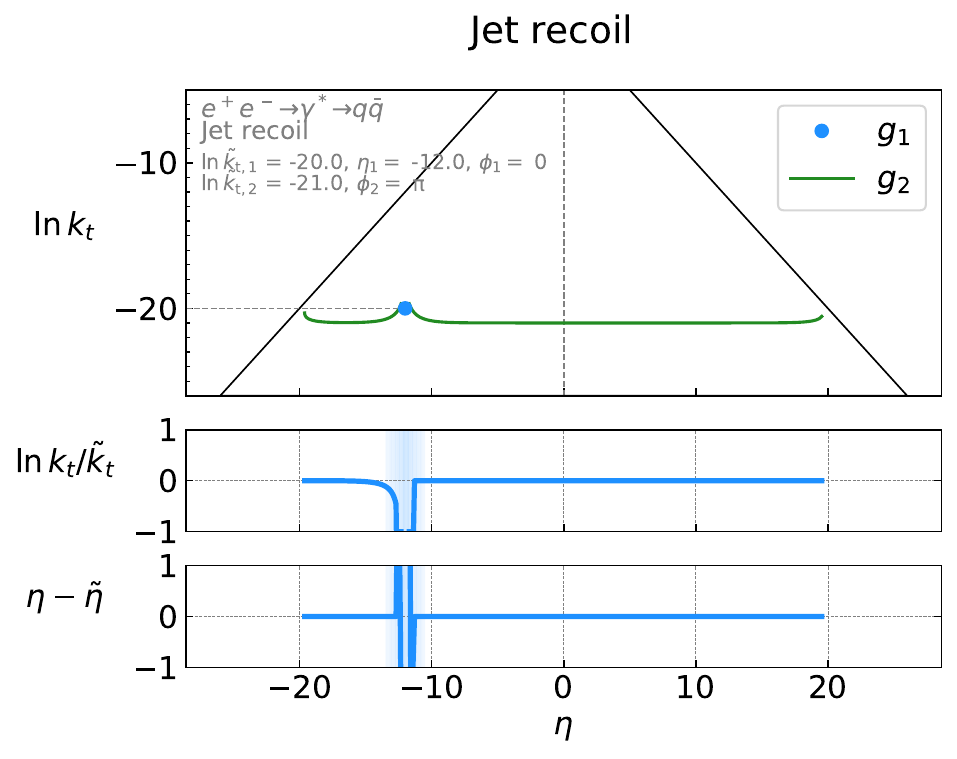}%
    \caption{Same as Fig.~\ref{fig:wrong-NLL-fo} for the jet-recoil attribution described in section~\ref{sec:example-as2}.}
    \label{fig:fo-tests-0}
\end{figure}

To illustrate that this recoil assignment does not lead to the spurious behaviour associated with purely dipole-local recoils, we consider the same fixed-order test as in Fig.~\ref{fig:wrong-NLL-fo}. This is shown in Fig.~\ref{fig:fo-tests-0}.
As can be observed in the two lower ratio plots, the kinematics of the first emission $g_1$ are only affected when the second emission, $g_2$, is at commensurate rapidity (which is relevant only at NNLL accuracy). 

Although the jet-recoil kinematic map uses the concept of angular ordering, we emphasise that it still reproduces the full pattern of soft emissions, an important feature of standard dipole showers related to their ability to correctly reproduce non-global resummations~\cite{Banfi:2006gy}.

\subsection{Determining $\calJ$}
\label{sec:jet-finding}
Having worked through several scenarios at $\mathcal{O}(\alpha_s^2)$ we are now ready to formulate our algorithm for determining $\calJ$ for an arbitrary partonic configuration. We will start by considering only gluon emissions, neglecting the splitting of gluons to quark pairs. The treatment of such splittings will be discussed in section~\ref{sec:edge-cases}.

The picture we gave at order $\mathcal{O}(\alpha_s^2)$ can be generalised using a clustering algorithm to define the recoiling jet. Indeed the regions identified in Fig.~\ref{fig:2-gluon-configs} were constructed according to angular ordering: one way of generalising this to all orders is through the Cambridge jet algorithm~\cite{Dokshitzer:1997in}, which clusters particles based on a purely angular pseudojet distance. 

\begin{enumerate}[label=(\arabic*)]
\item\label{config:primary} Let us start from a configuration where we have an arbitrary number of soft, strongly angular-ordered emissions from a $q\bar q$ system, $\lbrace g_1, \dots, g_n \rbrace$. We then perform an additional emission, $k$, from this configuration and seek to identify the recoiling jet, $\cal J$. In the instance where $k$ is also strongly ordered in angle, $\dots \ll \eta_i \ll \eta_k \ll \eta_{i+1} \ll \dots$, the same logic as in the previous section applies: the jet $\cal J$ should be composed of either the  quark or anti-quark, whichever is closest in angle to $k$, as  well as all particles which are closer in angle to the (anti-)quark than $k$ --- i.e.~all the particles that the emission $k$ does not resolve from the quark.
In the context of a clustering of this specific configuration with the Cambridge algorithm, this would correspond to all the particles previously clustered with the (anti-)quark.
For two emissions, this reduces to the situation described in the previous section for an emission in region 1 of Fig.~\ref{fig:2-gluon-configs} --- or regions 2 (4) if $k$ is the most collinear emission to the quark (anti-quark).

\item\label{config:secondary} For the same angular-ordered configuration, we now turn to the case where the new emission $k$ is more collinear to a previous emission, $g_i$, than to either of the Born partons. In this case, again, angular ordering dictates the recoiling jet $\cal J$ should include that particle, $g_i$, as well as any particles that would be clustered with $g_i$ before $k$ itself, within the Cambridge clustering sequence. At order $\mathcal{O}(\alpha_s^2)$, this corresponds to an emission in region 3 of Fig.~\ref{fig:2-gluon-configs}.

\end{enumerate}

For more complex configurations with nested splittings and without the constraint of strong angular ordering, the approach generalises; we determine the jet $\calJ$ by clustering the event, after the addition of an emission $k$, and stopping at the point in the sequence where $k$ would be clustered: $\cal J$ is simply taken to be the pseudojet that is about to cluster with $k$.

A useful way to visualise such a clustering sequence, which we will use below, is by constructing the Lund plane representation of the event~\cite{Dreyer:2018nbf}. In that context, leaves of the Lund plane correspond to branches in the clustering sequence: i.e.,~configurations described in~\ref{config:primary} correspond to emissions on the primary Lund plane, and configurations in~\ref{config:secondary} to emissions on the secondary Lund leaf belonging to $g_i$. Similarly, nested configurations correspond to tertiary leaves, quaternary leaves, and so on.

\begin{figure}
    \centering
    \includegraphics[width=0.5\linewidth,trim={0 0 31cm 0},clip]{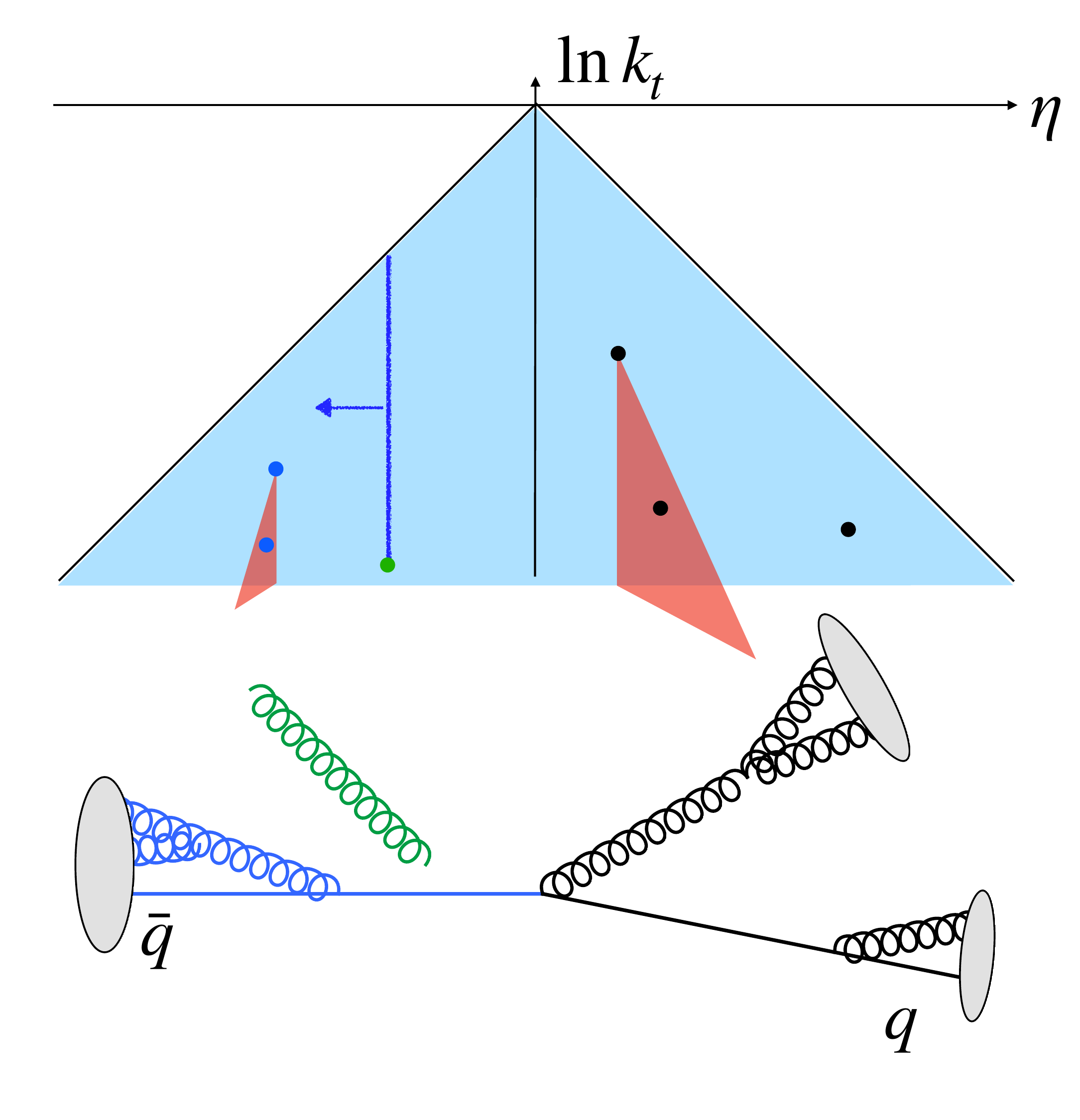}
    \caption{Lund plane and cartoon of event showing which partons are identified as $\calJ$. The new emission is shown in green and the partons comprising the jet in blue. The blue arrow on the Lund plane indicates the region of phase space in which particles will be included in $\calJ$.}
    \label{fig:EventClustering}
\end{figure}

The strategy described above is illustrated in Fig.~\ref{fig:EventClustering}, which shows an example event where the recoiling jet is highlighted in blue and the emission in green, along with the event's Lund plane representation.

Running a jet clustering algorithm every time an emission is added to the event, as previously discussed, would be computationally expensive. To avoid this, we adopt a simpler procedure that is equivalent in the limits relevant for NLL accuracy.
Such a procedure was formulated in Ref.~\cite{Hamilton:2020rcu}, with a focus on the correct assignment of colour factors in dipole and antenna parton showers. We adapt this strategy such that we can reconstruct the list of partons inside $\calJ$. Whereas the original algorithm kept track of the total flavour of $\calJ$, we need to have access to the full kinematics of particles that make up $\calJ$ (i.e.~to keep a list of the indices of those particles in the event record).

\begin{figure}
    \centering
    \includegraphics[width=0.5\linewidth,trim={0 0 30cm 0},clip]{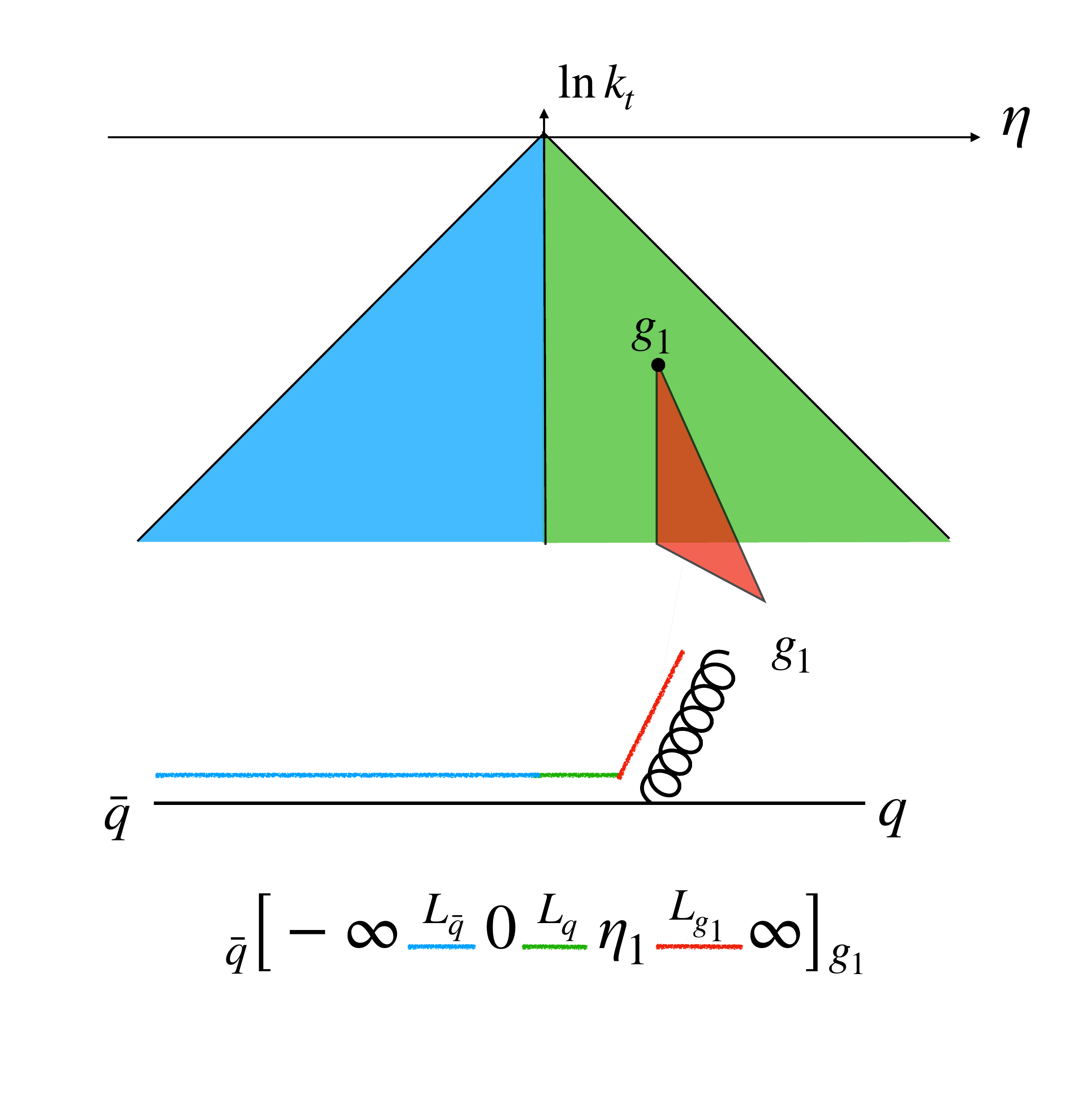}
    \caption{Illustration of the connection between intervals and regions of the Lund plane for the $\bar{q} g_1$ dipole in the above three-parton event. The coloured lines on the cartoon of the event correspond with the colour of the Lund leaf on which an emission appears, and thus which parton(s) in the diagram are considered as the emitter. }
    \label{fig:Intervals-example}
\end{figure}

During the shower evolution we construct, for each colour-anticolour dipole $(l, m)$, a list of rapidity intervals which point to a specific Lund leaf --- i.e.,~a branch of the would-be clustering sequence, denoted $L_p, L_q \dots$:
\begin{equation}
\leftindex_l{\Big[} -\infty \dots \eta_a \stackrel{ \mathlarger{L_p}}{\quad} \eta_b \stackrel{ \mathlarger{L_q}}{\quad} \eta_c \dots +\infty  \Big]_m \, .
\end{equation}
For bookkeeping purposes we always take the $\bar{3}$-end (the anticolour end) of the dipole to be at negative rapidities and the $3$-end (the colour end) at positive rapidities. The correspondence between intervals and regions of the Lund plane is illustrated in Fig.~\ref{fig:Intervals-example}.
Each Lund leaf $L_r$, with $r$ the index of the particle that ``owns" the leaf, is represented by a list of tuples:
\begin{align}
\begin{bmatrix} x &\quad y & \quad \dots \\ \eta_x &\quad  \eta_y & \quad \dots \end{bmatrix}_{L_r}
\end{align}
where the indices $x,y\dots$ refer to the particles that belong to the leaf, with rapidities $\eta_x, \eta_y, \dots$ with respect to the particle $r$.
The rapidity of $i$ with respect to $j$ is given by\footnote{
Here we only consider massless particles.}
\begin{equation}
    \eta_{ij} = -\ln \tan \frac{\theta_{ij}}{2} \,.
\end{equation}
Once an emission, $k$ from the $(\tilde l, \tilde m)$ dipole has been generated, we identify a rapidity $\eta_k$ as:
\begin{equation}
\eta_k = 
    \begin{cases}
        |\eta_{k \tilde m}| \ & \text{for} \ \ \eta_{k \tilde m} > \eta_{k \tilde l} \\
        -|\eta_{k \tilde l}|\  & \text{for} \ \ \eta_{k \tilde l} > \eta_{k \tilde m} ,
    \end{cases}\,.
\end{equation}
Specifically this is the rapidity (in the event frame) with respect to whichever end of the dipole the emission is closest to in angle, signed such that positive rapidities are collinear to the colour end of the dipole, while negative rapidities are collinear to the anticolour end.

We search for the interval in $\leftindex_l{[} \dots ]_m$ that contains $\eta_k$,  and thus identify the leaf $L_r$ on which the emission appears. 
The new emission is then added to $L_r$, and $\calJ$ deemed to consist of all particles on $L_r$ which have larger rapidities than the emission (i.e.,~particles which would be clustered with $r$ before the emission),
\begin{equation}
\calJ = \Big\{ i \in L_r \,\,\big|\,\, \eta_i > \eta_k \Big\}\,.
\end{equation}
This is equivalent to identifying $\cal J$ with the pseudojet that would be clustered with $k$, in the Cambridge algorithm.

\subsection{Update of the interval list}
\label{sec:interval-update}

Each time an emission is accepted, the list of intervals is updated (the updating of the Lund leaves was dealt with in the previous subsection). For updating the intervals we follow the same procedure as in Ref.~\cite{Hamilton:2020rcu};
for gluon emission $k$ from the $(l,m)$ dipole, this dipole splits into two dipoles $(l,k)$ and $(k,m)$, which necessitates splitting also the original list of intervals $\leftindex_l[\dots ]_m$. Consider an emission from the $(l,m)$ dipole which has intervals
\begin{equation}
\leftindex_l{\Big[} -\infty \dots \eta_a \stackrel{ \mathlarger{L_p}}{\quad} \eta_b \stackrel{ \mathlarger{L_q}}{\quad} \eta_c \dots +\infty  \Big]_m \,,
\end{equation}
with (say) $\eta_b < \eta_k < \eta_c$, i.e.~the emission belongs to leaf $L_q$. We split that list of intervals into two sets belonging to the two new dipoles,
\begin{multline}
    \leftindex_l{\Big[} -\infty \dots \eta_a \stackrel{ \mathlarger{L_p}}{\quad} \eta_b \stackrel{ \mathlarger{L_q}}{\quad} \max(0,\eta_k) \stackrel{ \mathlarger{L_k}}{\quad} +\infty \Big]_k \\
    \leftindex_k{\Big[} -\infty \stackrel{ \mathlarger{L_k}}{\quad} \min(0,\eta_k) \stackrel{ \mathlarger{L_q}}{\quad} \eta_c \dots +\infty  \Big]_m \,.
\end{multline}

To help clarify the procedure above, we provide a short worked-out example of how the algorithm functions, starting from a single $q \bar q$ dipole. In this case it is clear that if this dipole emits at positive (negative) rapidity then the emission will fall on the Lund leaf of the quark (anti-quark). The intervals list is therefore initialised as
\begin{equation}
\leftindex_{\bar{q}}{\Big[} -\infty \stackrel{ \mathlarger{L_{\bar q}}}{\quad} 0 \stackrel{ \mathlarger{L_q}}{\quad} +\infty  \Big]_q \,
\label{eq:qq-init-intervals}
\end{equation}
and the two Lund leaves initialised as empty lists. Consider that an emission $g_1$ is now generated with $\eta_1 > 0$. From the intervals list, the algorithm determines that the emission appears on the leaf of the quark, $L_q$, which is updated as follows:

\begin{align}
    \begin{bmatrix} \,\,\, \\ \,\,\, \end{bmatrix}_{L_q} \rightarrow \begin{bmatrix} g_1 \\ \eta_1 \end{bmatrix}_{L_q} \, .
\end{align}
A new empty leaf, $L_{g_1}$, is created for the gluon. As the $\bar q q$ dipole has now been split into two dipoles, $\bar q g_1$ and $g_1 q$ respectively, the intervals list must also be partitioned. The split occurs at $\eta_1$ with everything to the left (lower rapidity) becoming part of the $\bar q g_1$ dipole's interval list, and everything to the  right (larger rapidity) forming part of the $g_1 q$ intervals list. The newly created $\bar q g_1$ intervals list is given by
\begin{equation}
\textcolor{blue}{\leftindex_{\bar{q}}{\Big[} -\infty \stackrel{ \mathlarger{L_{\bar q}}}{\quad} 0 \stackrel{ \mathlarger{L_q}}{\quad}} \;\eta_1 \stackrel{ \mathlarger{L_{g_1}}}{\quad} +\infty  \Big]_{g_1} \,,
\end{equation}
where the part taken from the old $\bar q q$ interval is highlighted in blue and the boundary of the new interval pointing to $L_{g_1}$ is given by $\max(0,\eta_1) = \eta_1$ as the new interval corresponds to the colour end of the dipole. The $g_1 q$ intervals list is constructed similarly:
\begin{equation}
\leftindex_{g_1}{\Big[} -\infty \stackrel{ \mathlarger{L_{g_1}}}{\quad} 0\; \textcolor{blue}{\stackrel{ \mathlarger{L_q}}{\quad} +\infty  \Big]_q} \,,
\end{equation}
where the boundary of the new interval is $\min(0,\eta_1) = 0$ as it is at the anticolour end of the dipole.

The algorithm can then proceed in the same fashion for the next emission, starting from the intervals list corresponding to whichever dipole is now emitting.

\subsection{Special cases}
\label{sec:edge-cases}

In practice a few special cases arise, both in the identification of the recoiling jet, and in the treatment of gluon splittings to quark pairs. We explain how we handle these below.

\subsubsection{Identification of the spectator}

When a small-angle dipole emits a gluon $k$ at large angle, there is a region of azimuthal angles for which both dipole ends $(\tilde l, \tilde m)$ are identified as part of the recoiling jet $\calJ$ when we follow the procedure of section~\ref{sec:jet-finding}. This would leave us with no spectator to construct the map. In that case, we choose to remove from $\calJ$ the dipole end $\tilde l$ or $\tilde m$ that is furthest in angle from $k$, and identify that particle as the spectator instead. Ultimately, one may wish to consider alternative ways of identifying the spectator, e.g.~from the list of particles other than the dipole ends.

This procedure does not affect the logarithmic accuracy of the shower. To see this we note that for the emission to induce significant recoil in $\calJ$, it must be emitted with $k_t \sim k_{t\calJ}$. Furthermore, this situation only occurs in an $\mathcal{O}(1)$ rapidity-azimuth region. Any dangerous assignment of recoil therefore occurs only in an $\mathcal{O}(1)$ region of phase space and so the effect is suppressed by a power of $\alpha_s$ with no accompanying logarithms. As such the effect is beyond NLL accuracy.\footnote{We have verified numerically that the frequency at which these configurations occur agrees with this expectation.}

\subsubsection{Gluon splittings $g\to q\bar q$}

In the case where a gluon splits collinearly to a quark-antiquark pair, it is physically clear which particle is splitting --- namely the gluon. 
As this splitting channel only has a collinear singularity, the assignment of recoil for (suppressed) wide-angle splitting instead does not affect NLL accuracy. 
Let us consider that, algorithmically, the gluon is replaced with (say) the anti-quark in the event record and that the emission is identified as the quark. We choose to identify the new anti-quark as the single particle comprising the recoiling jet $\calJ$. The spectator is then simply taken to be the other end of the emitting dipole. 

\begin{figure}[h]
    \centering
    \includegraphics[width=0.4\linewidth,trim={0 0 30cm 0},clip]{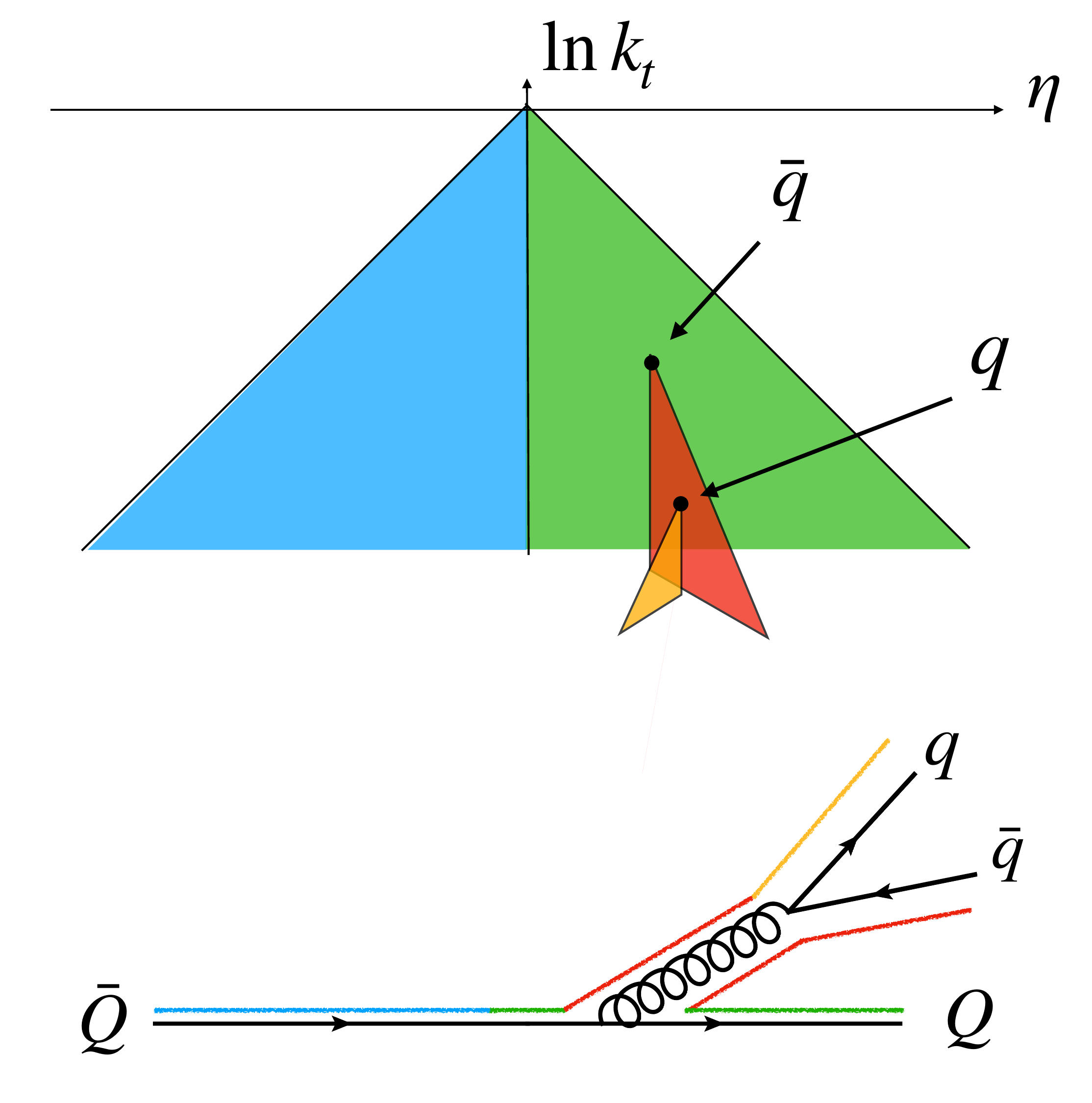}
    \caption{Illustration of the Lund plane structure of an event with a $g\rightarrow q\bar{q}$ splitting. In this example the gluon has been replaced by an anti-quark, and a quark was emitted. The coloured lines show which particles each Lund leaf corresponds to. Note that the leaf labelled as $\bar{q}$ would have belonged to the gluon up until it split to the $\bar{q} q$ pair. Emissions at smaller rapidities than $\eta_q$ on this leaf give transverse recoil to $\calJ=\{\bar{q},q \}$, as if they had been emitted from the gluon before it split.}
    \label{fig:gqqbar_plane}
\end{figure}

For $g\rightarrow q\bar q$ splitting there are two lists of intervals that need updating, corresponding to the two dipoles connected to the splitting gluon, denoted as $g$. We continue  with the example where the gluon is replaced with an anti-quark $\bar q$, and the emission is a quark $q$. We calculate $\eta_k = \eta_{q\bar q}$ as before, and update the interval lists for the two affected dipoles as
\begin{align}
\leftindex_l{\Big[} -\infty \dots \eta_b \stackrel{ \mathlarger{L_g}}{\quad}+\infty  \Big]_g &\longrightarrow \leftindex_l{\Big[} -\infty \dots \eta_b \stackrel{ \mathlarger{L_{\bar{q}} }}{\quad} \eta_{q \bar q} \stackrel{\mathlarger{L_{q} }}{\quad} +\infty  \Big]_q \,, \label{eq:qqbar-3end}\\
\leftindex_g{\Big[} -\infty \stackrel{ \mathlarger{L_g}}{\quad} \eta_c \dots +\infty  \Big]_m & \longrightarrow \leftindex_{\bar q}{\Big[} -\infty \stackrel{ \mathlarger{L_{\bar q} }}{\quad} \eta_c \dots +\infty  \Big]_m \,.\label{eq:qqbar-3barend}
\end{align}
The original Lund leaf $L_g$ becomes $L_{\bar q}$ (algorithmically they are owned by the same particle, since $g$ became $\bar q$), and $q$ appears as an emission on that leaf. As with gluon emission, a new leaf, $L_{q}$, is created for the quark (see Fig.~\ref{fig:gqqbar_plane}).

It can happen that the gluon splits to a large-angle $q\bar q$ pair, i.e.~that $\eta_{q\bar q} < \eta_b$ in the example above. In that case  (which contributes beyond NLL accuracy) we choose to replace Eq.~(\ref{eq:qqbar-3end}) with
\begin{align}
    \leftindex_l{\Big[} -\infty \dots \eta_b \stackrel{ \mathlarger{L_g}}{\quad}+\infty  \Big]_g &\longrightarrow \leftindex_l{\Big[} -\infty \dots \eta_b \stackrel{\mathlarger{L_{q} }}{\quad} +\infty  \Big]_q \,.
\end{align}
The case where the gluon becomes a quark is treated similarly.

\subsection{Born gluons}
\label{sec:alg-h2gg}

The same algorithm can be applied to the case of cyclically colour-connected dipoles in the Born state, such as in the $e^+e^- \to H \to gg$ process. The only adaptation needed is in the initialisation of the list of intervals for the Born event (i.e.~two gluons $g_a$ and $g_b$), which is taken to be:
\begin{align}
&\leftindex_{g_a}{\Big[} -\infty \stackrel{ \mathlarger{L_{g_a}}}{\quad} 0 \stackrel{ \mathlarger{L_{g_b}}}{\quad} +\infty  \Big]_{g_b} \,, \\
&\leftindex_{g_b}{\Big[} -\infty \stackrel{ \mathlarger{L_{g_b}}}{\quad} 0 \stackrel{ \mathlarger{L_{g_a}}}{\quad} +\infty  \Big]_{g_a} \,,
\end{align}
instead of Eq.~(\ref{eq:qq-init-intervals}). This can be generalised to more complex Born configurations. The update of the list after each emission proceeds as in section~\ref{sec:interval-update}. A representative subset of the fixed- and all-order tests performed in section~\ref{sec:num-tests} is repeated for $H\to gg$ in Appendix~\ref{sec:app-h2gg}.

\section{Numerical tests}
\label{sec:num-tests}

We now turn to results of numerical tests of the jet recoil scheme. To perform these, we implemented the map of section~\ref{sec:shower-desc} as a user-defined shower within the PanScales public code, v.0.2.0~\cite{vanBeekveld:2023ivn}. We confirm the validity of the approach at fixed order (sections~\ref{sec:fo-tests-alg} and~\ref{sec:fo-tests-NLL}) and all orders (section~\ref{sec:nll-tests}). For the latter, we use the logarithmic accuracy tests available in the PanScales public code. In both cases the jet-recoil shower uses a transverse evolution variable $v=k_t$ as defined in Eqs.~(\ref{eq:dipole-map-pk}),~(\ref{eq:kperp}), leading-order DGLAP splitting kernels for the emission density and the strong coupling is evaluated with 2-loop running in the CMW scheme~\cite{Catani:1990rr}.

\subsection{Tests of the algorithm}
\label{sec:fo-tests-alg}

First, we establish that the implementation of the jet recoil procedure reproduces the expected identification of $\calJ$, as described in section~\ref{sec:shower-desc}. To begin with, a Born $q\bar q$ event is generated, to which we add a number $n$ of emissions with fixed kinematics, $\lbrace g_1, \dots, g_n \rbrace$. From this fixed event, a further emission is performed, $g_{n+1}$ and its transverse momentum and rapidity are scanned over (in these tests we keep the generation-level azimuth $\phi_{n+1}$ fixed). We let the shower find the recoiling jet $\cal J$, as per the algorithm of section~\ref{sec:jet-finding}; finally we perform a Lund declustering of the whole event, using the Cambridge algorithm, and examine which particles make up the jet $\cal J$ as a function of the Lund coordinates of the last emission.

\begin{figure}[ht!]
    \centering
    \includegraphics[width=.49\textwidth]{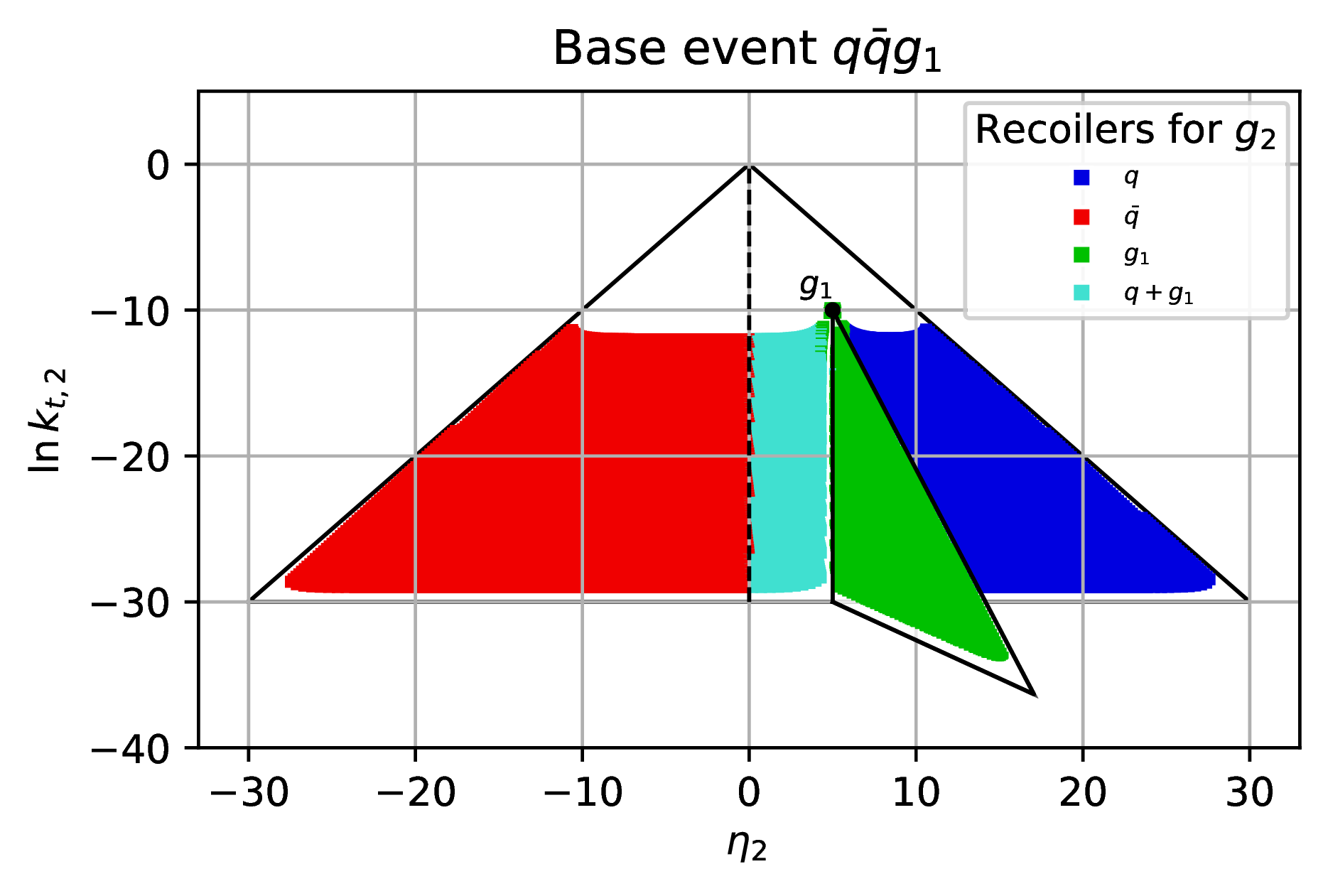}%
    \includegraphics[width=.49\textwidth]{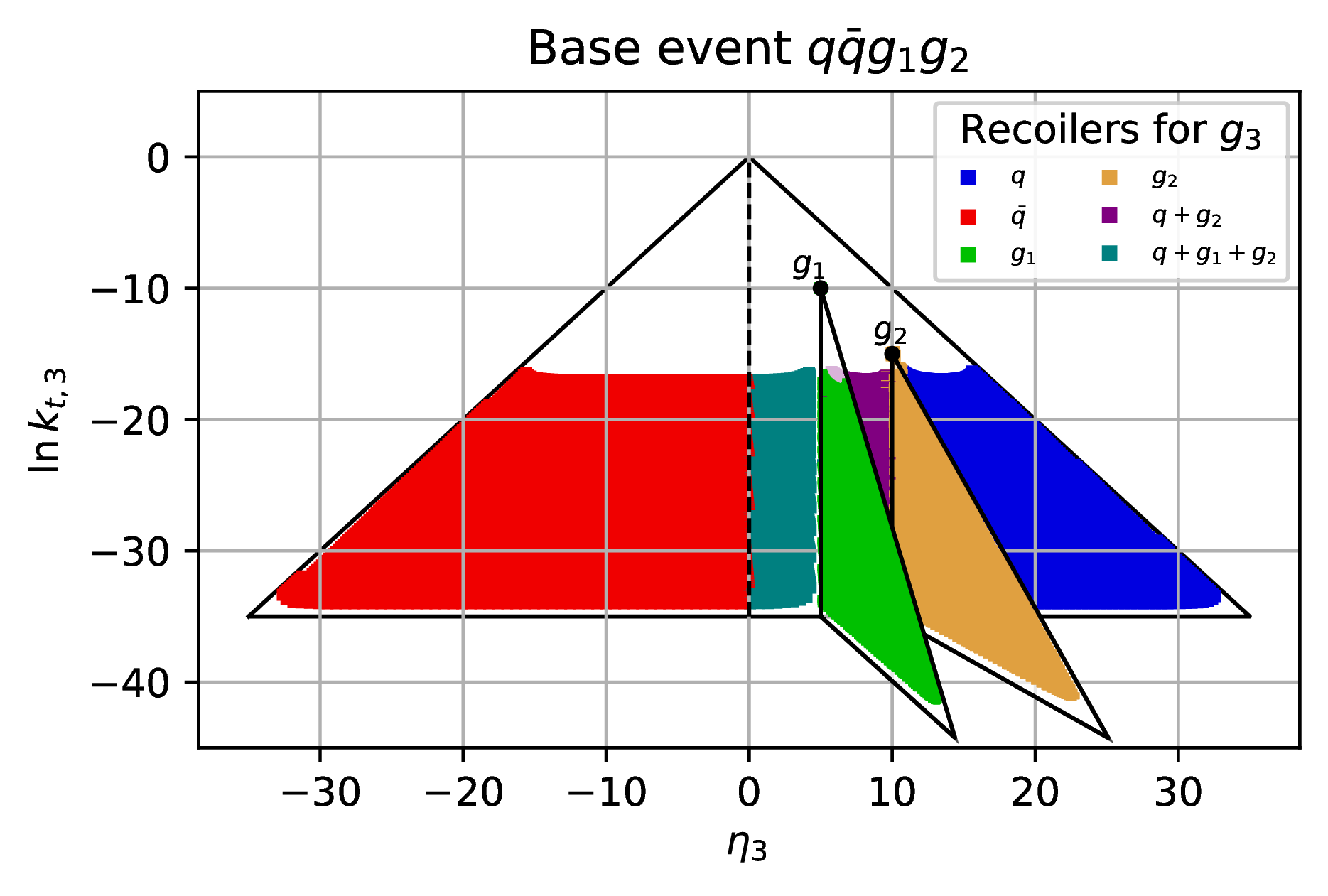}\\
    \includegraphics[width=.49\textwidth]{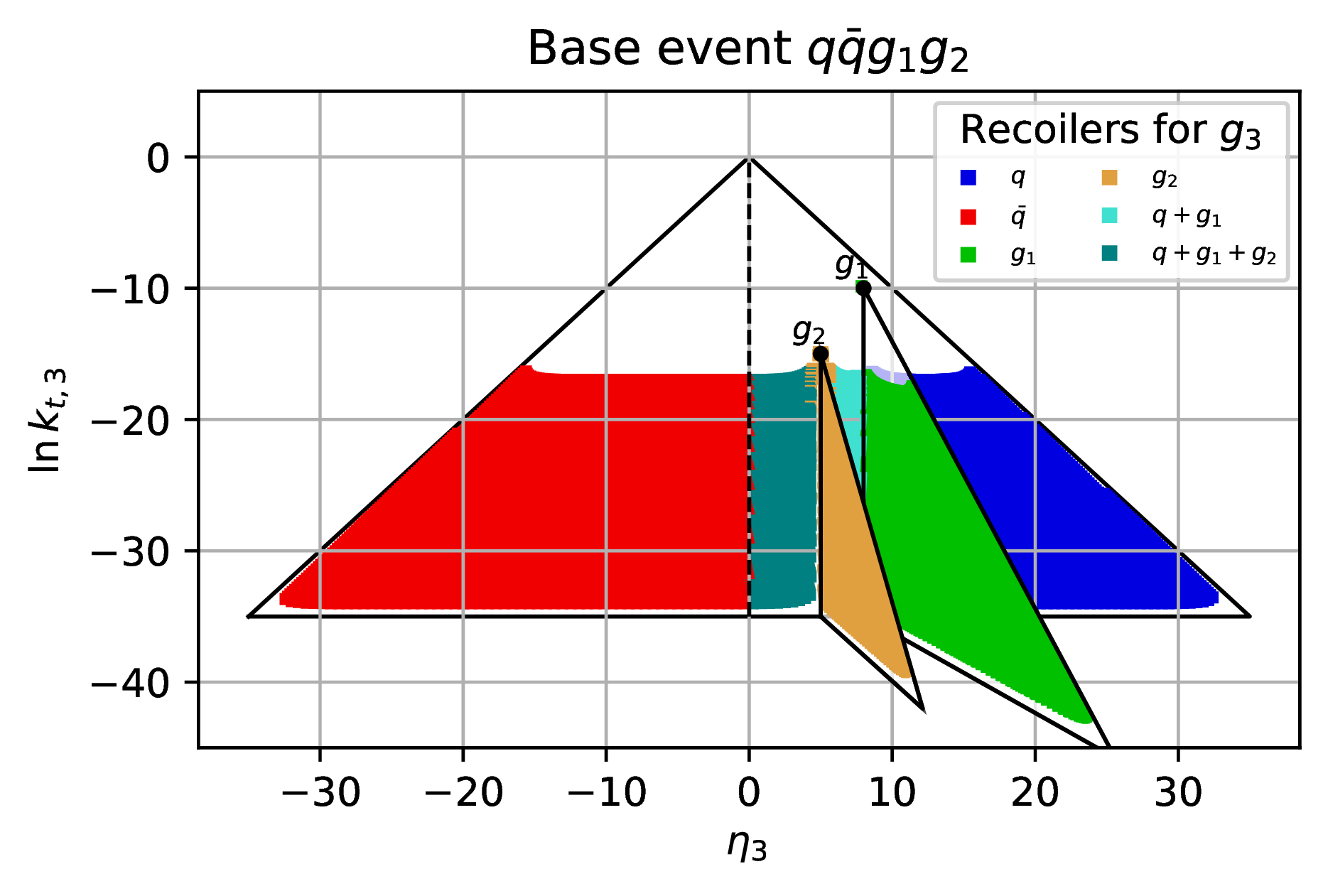}
    \includegraphics[width=.49\textwidth]{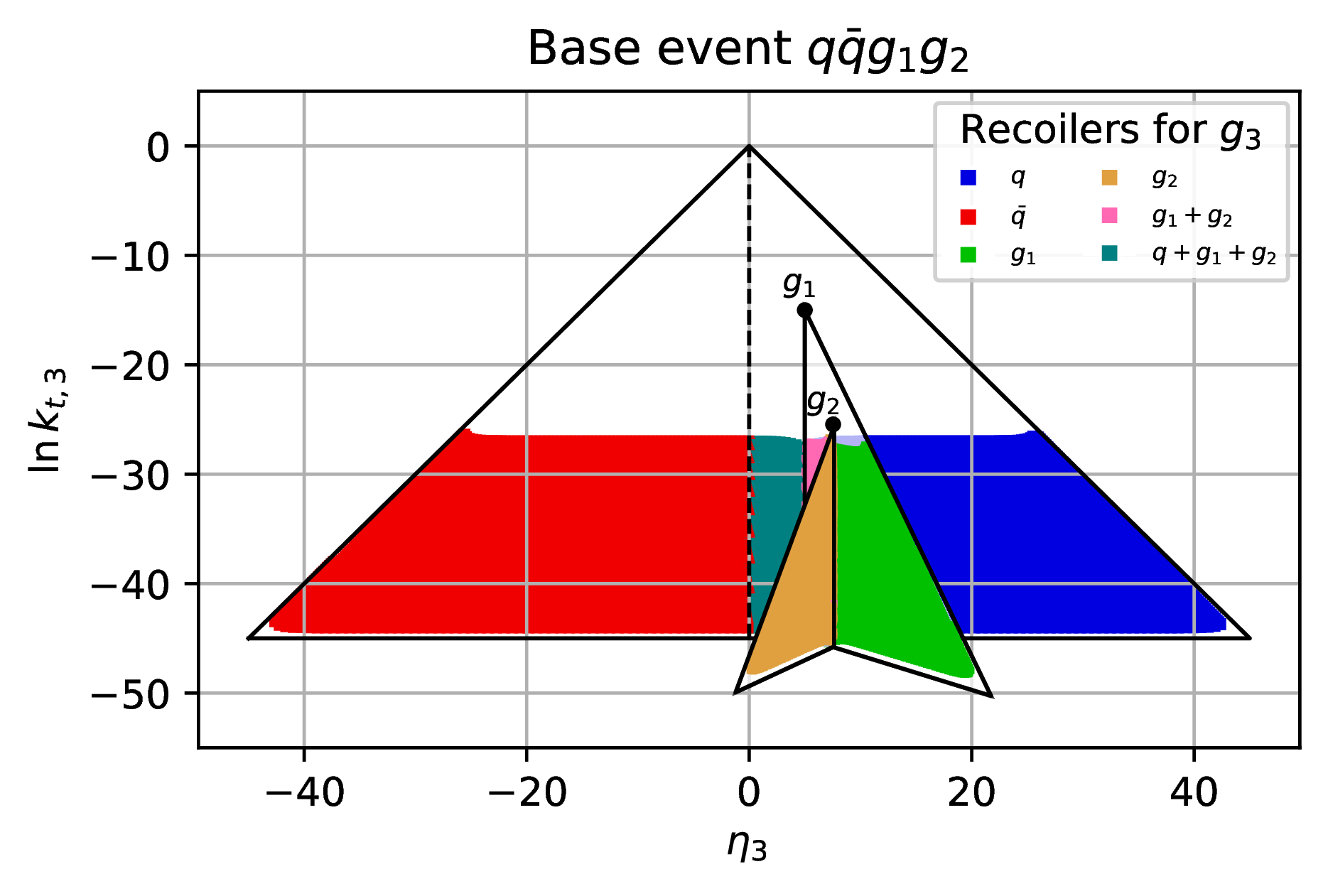}
    \caption{Fixed-order tests of the implementation of the jet-recoil shower, at second order for a single primary (top left), and at third order for two primaries (top right, bottom left) with different hierarchies of $g_1$ and $g_2$, and for a tertiary configuration (bottom right). The composition of the recoiling jet $\calJ$ is shown in the legend as a colour list.}
    \label{fig:set-recoilers}
\end{figure}

The composition of the recoiling jet for the emission of $g_{n+1}$ is shown in Fig.~\ref{fig:set-recoilers}, for various starting configurations. We show only a few of the less trivial configurations, arising from emissions in the same hemisphere, with different orderings of the emissions in the base event. Coloured points are located at the coordinates of the last emission, $(\ln k_{t,n+1}, \eta_{n+1})$, and the colour corresponds to a given composition of the jet $\cal J$. The identification of the jet is in agreement with the underlying argument of angular ordering, for all configurations, in the relevant logarithmically-enhanced regions. In cases where all emissions are well separated in rapidity, the identification of the recoiling jet is unambiguous. When the last emission, $g_{n+1}$, is at angles commensurate with another parton in the event, $\eta_{n+1}\simeq \eta_i$, the recoiling jet may not be identical to that which would be found with the Cambridge algorithm (as indicated by the smatterings of wrongly-coloured points close to the top of the secondary leaves). As explained in section~\ref{sec:jet-finding}, this is only relevant beyond NLL.

\subsection{Fixed-order tests of NLL configurations}
\label{sec:fo-tests-NLL}

Next we verify that our scheme fulfils the factorisation properties (the \emph{fixed-order} criterion) laid out in Refs.~\cite{Dasgupta:2018nvj,Dasgupta:2020fwr}, which are a necessary condition for NLL accuracy. Namely, a parton shower should generate the correct IR effective matrix element for any pair of emissions that are widely separated in at least one logarithmic variable (i.e.~transverse momentum, rapidity or some linear combination thereof). We verify this by following the fixed-order testing procedure established in Refs.~\cite{Dasgupta:2018nvj,vanBeekveld:2022zhl,vanBeekveld:2023chs}.

Typically this condition is fulfilled in standard dipole showers when partons are strongly ordered in transverse momenta, but not when they are emitted at commensurate $k_t$ and widely-separated rapidities, as explained in section~\ref{sec:recoil-log-accuracy}.
Below we perform fixed-order tests of the jet recoil scheme (similarly to Figs.~\ref{fig:wrong-NLL-fo} and~\ref{fig:fo-tests-0}), whereby we verify that 
for a wide range of given starting configurations, the last shower emission does not significantly modify the kinematics of other emissions in the event in the limit where the scale hierarchies become large (in one or the other direction of the Lund plane).

 We again start from a fixed kinematic configuration ($q\bar q g_1$, or $q\bar q g_1 g_2$), and generate an extra gluon emission $g$ with a value of the transverse momentum $k_t$ commensurate with that of the previous emission, and scan over the longitudinal variable. As before we then examine the change in the Lund coordinates of the previous emissions.
For NLL accuracy the last emission is only allowed to significantly impact the kinematics of other emissions when it is close in both logarithmic variables (these configurations contribute at NNLL).

\begin{figure}[ht!]
    \centering
    \includegraphics[height=5.6cm,page=3,trim={0 0 0 1cm}, clip]{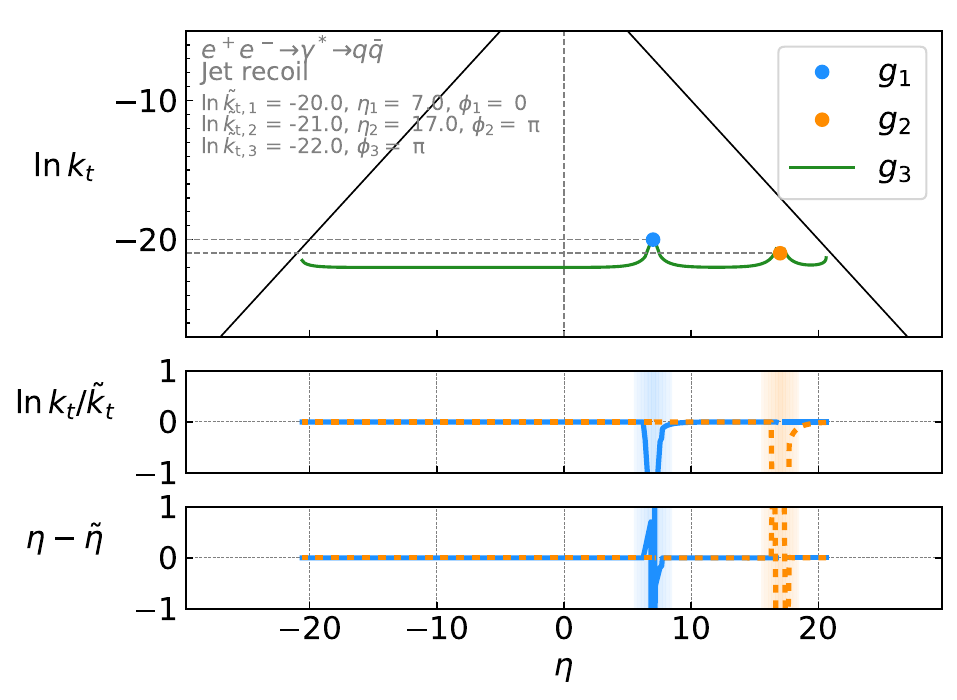}%
    \includegraphics[height=5.6cm,page=2,trim={0 0 0 1cm}, clip]{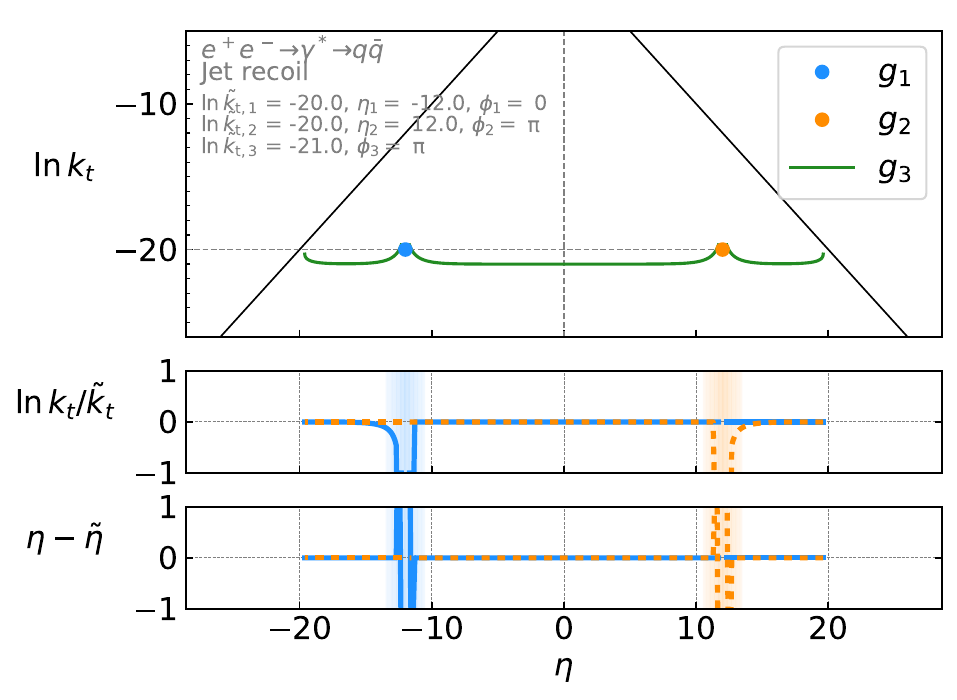}
    \caption{Fixed-order tests as in Figs.~\ref{fig:wrong-NLL-fo},~\ref{fig:fo-tests-0} of the jet recoil shower, with two primary gluon emissions. The contour of fixed shower ordering variable for the last emission is depicted in green on the main plot. The change in transverse momentum ($\ln k_t/\tilde{k_t}$, first ratio plot) and rapidity ($\eta - \tilde \eta$, second ratio plot) of the gluons before and after the last emission are shown in blue for $g_1$ (dashed orange for $g_2$), for an emission on the primary Lund plane.}
    \label{fig:fo-tests-1}
\end{figure}

\begin{figure}[ht!]
    \centering
    \includegraphics[height=5.6cm,page=5,trim={0 0 0 1cm}, clip]{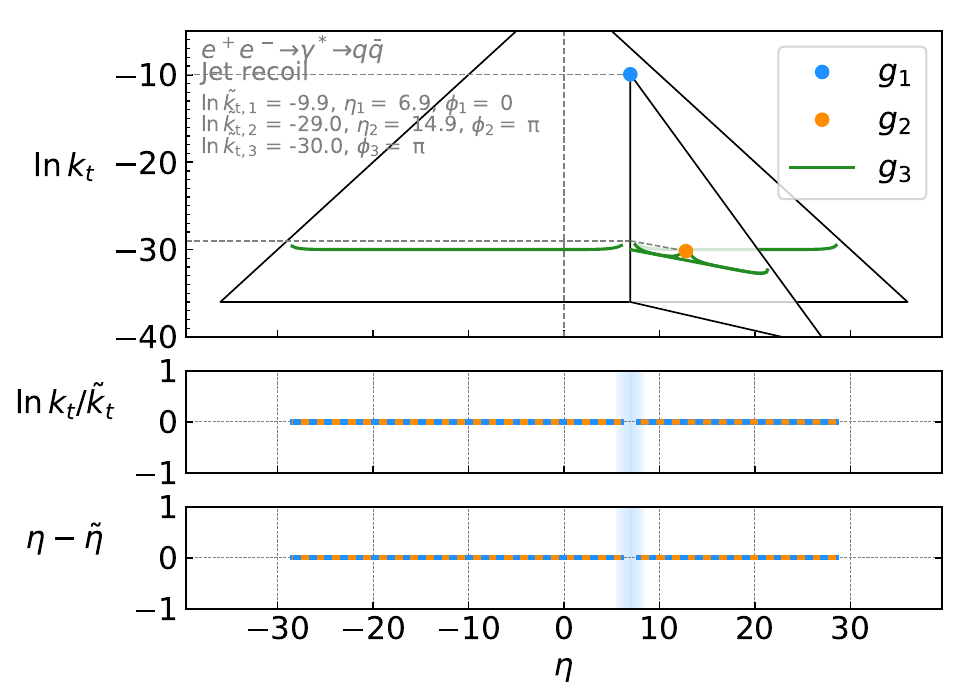}%
    \includegraphics[height=5.6cm,page=6,trim={0 0 0 1cm}, clip]{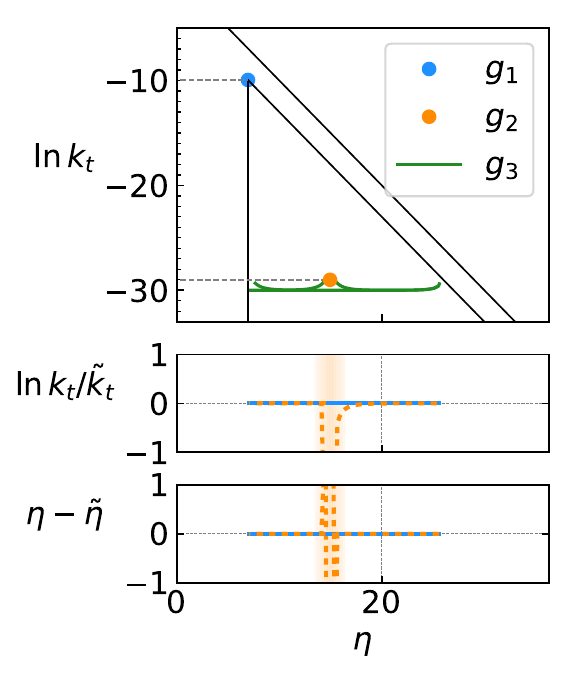}
    \caption{Same as Fig.~\ref{fig:fo-tests-1} for one primary and one secondary gluon: the primary plane is shown on the left, and the secondary plane belonging to gluon $g_1$ on the right. In the ratio plots, the change in $k_t$ or $\eta$ is shown for an emission on the primary Lund plane (left), and on the secondary leaf of $g_1$ (right).}
    \label{fig:fo-tests-sec}
\end{figure}

Fixed-order tests for $q\bar{q}g_1 +g$ were shown in Fig.~\ref{fig:fo-tests-0}. We also tested several configurations at higher order that lead to different recoiling jets. Fig.~\ref{fig:fo-tests-1} (Fig.~\ref{fig:fo-tests-sec}) illustrates such a test for a base event consisting of two primary gluons (respectively one primary, one secondary). In these tests, the jet recoil scheme fulfils the fixed order criterion for NLL accuracy. This gives confidence that the algorithm is functioning as intended at higher order. The all-order logarithmic structure is tested in the next section.

\subsection{All-order NLL tests}
\label{sec:nll-tests}

We establish the all-order accuracy of the jet recoil map, by comparing the shower predictions to known resummation results for a wide range of observables, as pioneered in Ref.~\cite{Dasgupta:2020fwr}.
Parton-shower predictions necessarily contain uncontrolled terms beyond the shower target accuracy. To isolate the NLL contribution, we use the logarithmic accuracy tests available in the PanScales public code~\cite{vanBeekveld:2023ivn}. This involves generating parton-shower predictions for each observable at several small $\alpha_s$ values, whilst keeping $\lambda = \alpha_s L$ constant. For each run one constructs
\begin{equation}
\delta^{\mathrm{NLL}} = \frac{\Sigma^{\mathrm{shower}}(\alpha_s,L)}{\Sigma^{\mathrm{NLL}}(\alpha_s,L)}-1
\end{equation}
where $\Sigma(\alpha_s,L)$ is defined in Eq.~\eqref{eq:sigmaNLL}.
If the shower is NLL accurate, $\delta^{\mathrm{NLL}}$ will vanish in the limit of $\alpha_s\to 0 $. Running parton showers in this limit ($\alpha_s \to 0, L \to \infty$) is numerically extremely challenging, and the efficient techniques developed within PanScales to mitigate these difficulties were indispensable in performing these tests.

As in Ref.~\cite{Dasgupta:2020fwr} we consider several global event-shape observables that cover a range of scalings with respect to single (soft-collinear) emission, $V\propto \frac{k_t}{Q} e^{-\betaobs |\eta|}$,
and vary in their sensitivity to emissions in different parts of phase space~\cite{Banfi:2004yd}. We also test the transverse momentum in a rapidity slice (a non-global observable) at single-logarithmic level~\cite{Dasgupta:2001sh,Dasgupta:2002bw}.
 Finally, we examine the subjet multiplicity, at next-to-double-logarithmic (NDL) level~\cite{Catani:1991pm, Medves:2022ccw}.
 The complete list of observables is given here:
\begin{itemize}
    \item the total and wide-jet broadenings $B_T$, $B_W$~\cite{Catani:1992jc,Catani:1992ua}, and the Cambridge jet resolution parameter $\sqrt{y_{23}}$~\cite{Dokshitzer:1997in} (all $\betaobs = 0$),
    \item the thrust $1-T$~\cite{Brandt:1964sa,Farhi:1977sg} ($\betaobs = 1$),
    \item the sum $S_{\betaobs}$ and max $M_{\betaobs}$ across primary Lund declusterings $\mathbb{P}$~\cite{Dreyer:2018nbf,Dasgupta:2018nvj,Dasgupta:2020fwr}, with values of the parameter $\betaobs \in \lbrace 0,\frac{1}{2},1\rbrace$,
    \begin{equation}
        S_{\betaobs} = \sum_{i\in \mathbb{P}} \frac{k_{t,i}}{Q} e^{-\betaobs |\eta_i|}\,,\quad M_{\betaobs} = \max_{i\in \mathbb{P}} \frac{k_{t,i}}{Q}e^{-\betaobs |\eta_i|}\,,
    \end{equation}
    \item the fractional moments of energy-energy correlations $\mathrm{FC}_{1-\betaobs}$~\cite{Banfi:2004yd},
    \item the transverse scalar sum in a rapidity slice~\cite{Dasgupta:2002bw}, \begin{equation}\frac{1}{Q}\sum_i p_{t,i} \, \Theta(\eta_{\max} - |\eta_i|)\,,\end{equation} with $\eta_{\max}=1$,
    \item the subjet multiplicity $N_\mathrm{subjet}$~\cite{Catani:1991pm} for Durham ($k_t$) jets~\cite{Catani:1991hj}.
    
\end{itemize}
One class of observables that are absent from our tests are those sensitive to spin correlations. While this is strictly part of general NLL accuracy, there exist several algorithms~\cite{Collins:1987cp,Knowles:1988hu,Hoche:2025anb} that can be integrated with parton showers to incorporate these effects~\cite{Richardson:2001df,Richardson:2018pvo,Karlberg:2021kwr,Hamilton:2021dyz}. Such effects can be safely neglected for observables that average over azimuthal modulations, like those we consider here. We defer the treatment of spin correlations with a jet-recoil map to future work.

To extract $\delta(0):=\lim_{\alpha_s \to 0} \delta^{\mathrm{NLL}}$, we generate parton-shower predictions for $\alpha_s \in [0.0224,0.0168,0.0112,0.0084,0.0056]$, keeping $\lambda= \alpha_s L = -0.42$ fixed. We perform two fits to carry out the $\alpha_s \to 0$ extrapolation: one with a quadratic function through the five $\alpha_s$ points, and one linear fit using only the three smallest $\alpha_s$ points. Our final central value for $\delta(0)$ is taken from the quadratic fit. The total uncertainty on $\delta(0)$ is taken to be the sum in quadrature of the statistical uncertainty from the quadratic fit, and a systematic uncertainty from the fit procedure. The systematic uncertainty is defined as the absolute value of the difference between the $\delta(0)$ values given by the quadratic and linear fits. 

For the subjet multiplicity, which does not exponentiate, we check that the shower correctly reproduces the NDL result. To this end we hold $\alpha_s L^2 = 5$ constant and produce predictions for $\alpha_s \in [0.003125,0.002,0.00128,0.00078125,0.0005]$. We then extract $\delta^{\mathrm{NDL}} = (N(\alpha_s,L)-N^{\mathrm{shower}}(\alpha_s,L))/\sqrt{\alpha_s}$ in the limit of $\alpha_s\to0$, using the same procedure as for the other observables.

\begin{figure}[h]
    \centering
    \includegraphics[width=0.7\linewidth, clip]{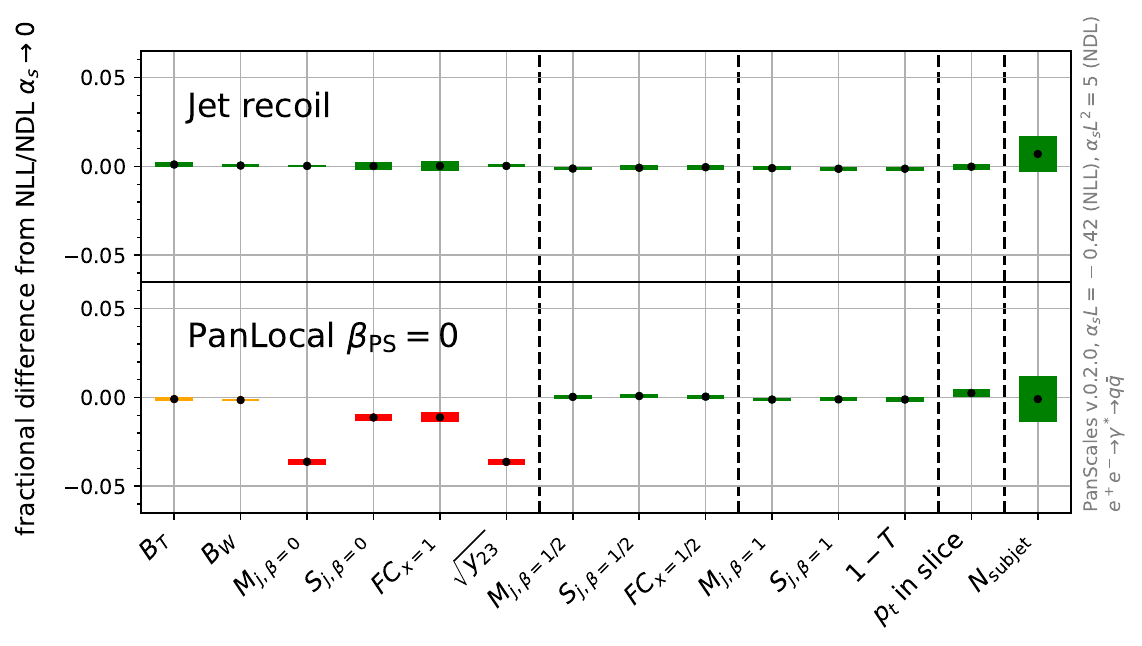}
    \caption{Results of NLL tests using our new jet recoil map (top), and for PanLocal with $\betaps = 0$ (bottom) --- which is NLL accurate only for $\betaps > 0$, for comparison. Green signals agreement with analytically resummed results to within $2\sigma$. Red denotes a $4\sigma$ or greater disagreement and orange denotes less than $4\sigma$ deviation from the resummed result for cases where the shower is known not to reproduce the correct resummed result.} 
    \label{fig:NLL_results}
\end{figure}

The final results are shown in Fig.~\ref{fig:NLL_results}, where consistency with zero indicates agreement with the analytic resummation. We also perform the same tests using the PanLocal shower (which achieves NLL accuracy only for $\betaps > 0$) setting $\betaps = 0$. We deliberately break NLL accuracy for PanLocal in order to demonstrate that the tests we perform here are capable of resolving the discrepancy. For the PanLocal ($\betaps = 0$) shower, we only expect a discrepancy in $\betaobs=0$ observables (the six leftmost).\footnote{In the case of the jet broadenings, $B_T$ and $B_W$, the observed departure from NLL is below $4\sigma$, but we expect the discrepancy to be small for these observables~\cite{Dasgupta:2020fwr}.} In the upper panel, we observe perfect agreement between the jet-recoil shower prediction and the exact NLL resummation for all the observables we consider. This constitutes a numerical proof that the jet recoil scheme can be used to achieve NLL accuracy in $k_t$-ordered dipole showers.

\section{Conclusions}
\label{sec:conclusions}

We have presented a new recoil scheme that can be used to achieve NLL accuracy in final-state, $k_t$-ordered dipole showers. In this scheme, recoil is distributed to ``jets" of partons in the event in a way that is closely related to angular ordering. The jets are constructed by a similar procedure as the one introduced in Ref.~\cite{Hamilton:2020rcu}; whereas the original algorithm focused on the correct assignment of colour factors at NLL level within the PanScales showers, we adapt this strategy to dynamically keep track of a list of particles associated with each emitting colour dipole. This circumvents the need to cluster particles at each stage in the shower evolution and is equivalent to Cambridge clustering for configurations where emissions are well-separated in rapidity.

We implemented a jet-recoil shower map as a user-defined shower within the PanScales public code, v.0.2.0~\cite{vanBeekveld:2023ivn}. We validated the NLL accuracy of the proposed map according to the criteria laid out in Ref.~\cite{Dasgupta:2020fwr}: these criteria require tests of the effective shower matrix element at fixed order, as well as logarithmic tests of the shower resummation for a wide variety of observables. For the latter, the numerical techniques developed within PanScales~\cite{Dasgupta:2020fwr, vanBeekveld:2022ukn} and implemented in the public code were indispensable.

Having established that giving recoil to jets of particles is a viable path to NLL accuracy, we envisage that this procedure could be used with existing dipole-local kinematic maps to achieve NLL accuracy in the showers that rely on them whilst preserving many of their existing features. Application of the jet recoil paradigm to antenna-type showers --- such as \Vincia~\cite{Fischer:2016vfv,Brooks:2020upa}, which share transverse recoil between both ends of the dipole, is left to be investigated in future work.

\section*{Acknowledgements}
We are particularly grateful to Gavin Salam and Gregory Soyez for discussions relating to their unpublished work~\cite{SalamSoyez:unpublished} and  for helpful comments on the manuscript. We also acknowledge the pilot project done by Wilton Deany for his honours thesis at Monash~\cite{WiltonThesis}. 
We wish to thank the PanScales collaboration as a whole, in particular for their work on the public code framework which makes logarithmic tests of parton showers possible. 
We also thank Basem El-Menoufi for useful discussions in the early stages of the project, as well as Riley Henderson for comments on the manuscript. This work was supported by the Australian Research Council under Discovery Project grants DP230103014 (JH, PS), DP220103512 (PS) and Discovery Early Career Researcher Award DE230100867 (LS).

\appendix
\section{Results for $H\to gg$}
\label{sec:app-h2gg}

Here, we collect the results of fixed- and all-order tests of the jet-recoil scheme applied to the $H\to gg$ process. We repeat that once the list of intervals is initialised (see section~\ref{sec:alg-h2gg}), the algorithm proceeds as in the case of a $q\bar q$ dipole.

\begin{figure}[ht!]
    \centering
    \includegraphics[width=.49\textwidth]{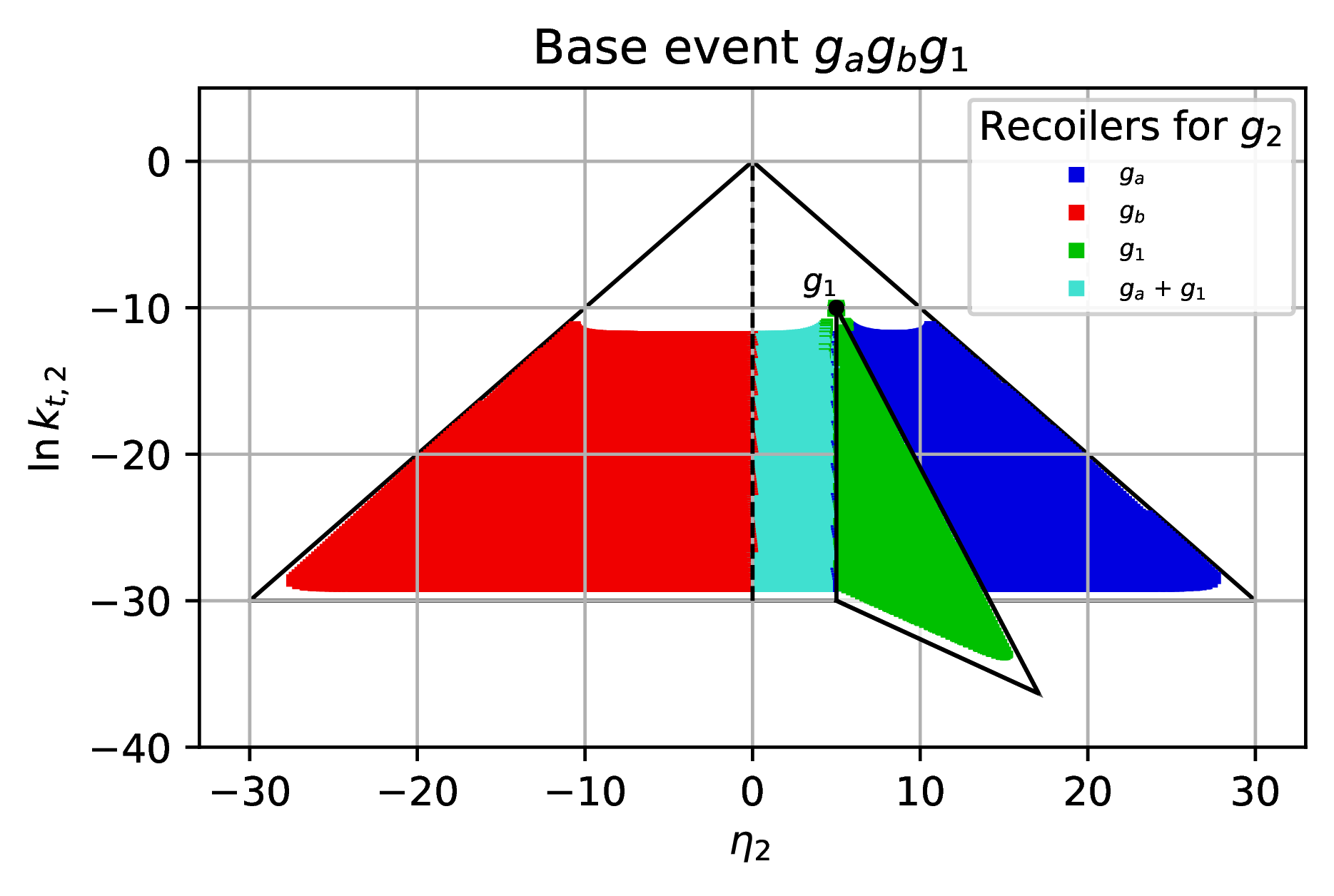}%
    \includegraphics[width=.49\textwidth]{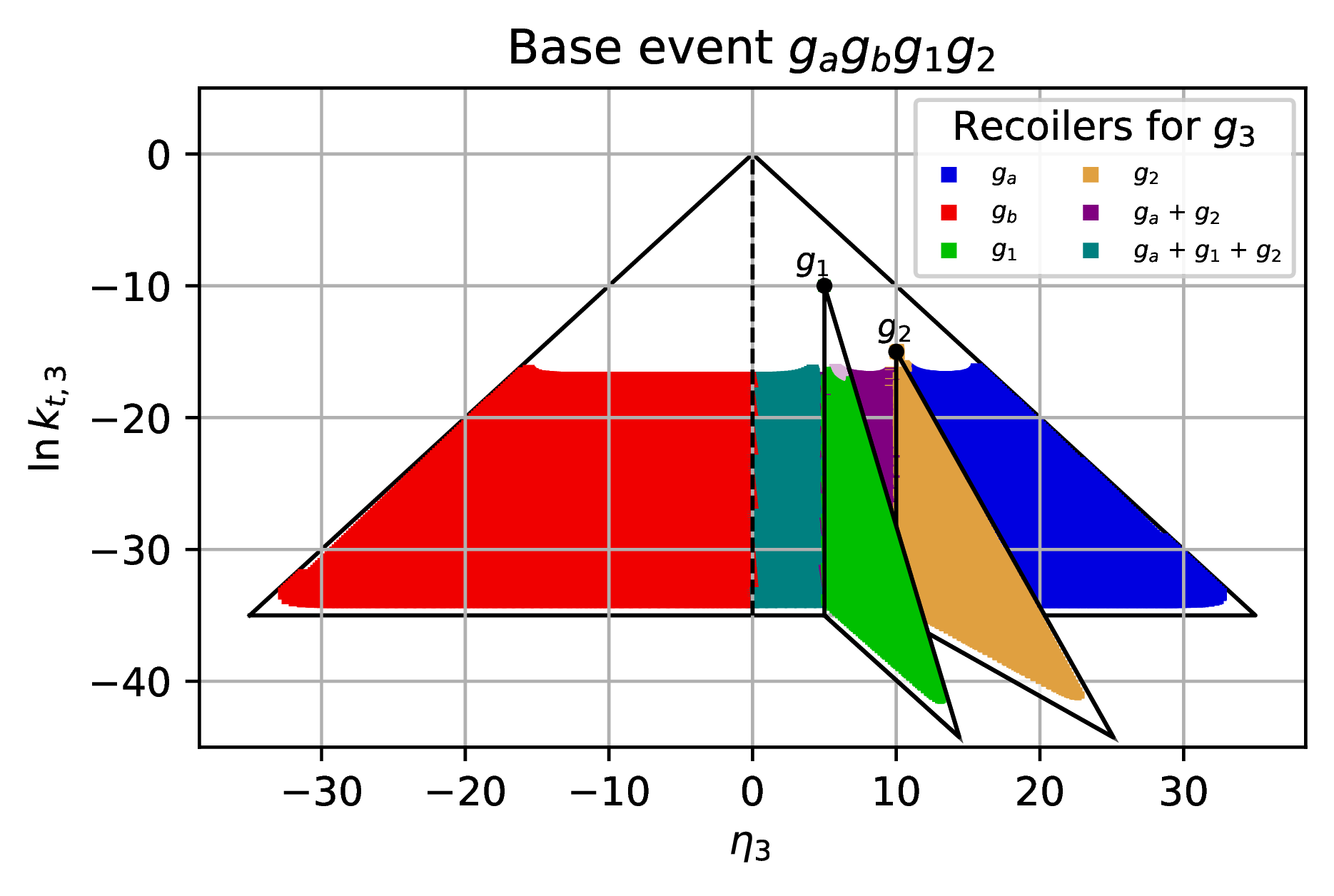}\\
    \includegraphics[width=.49\textwidth]{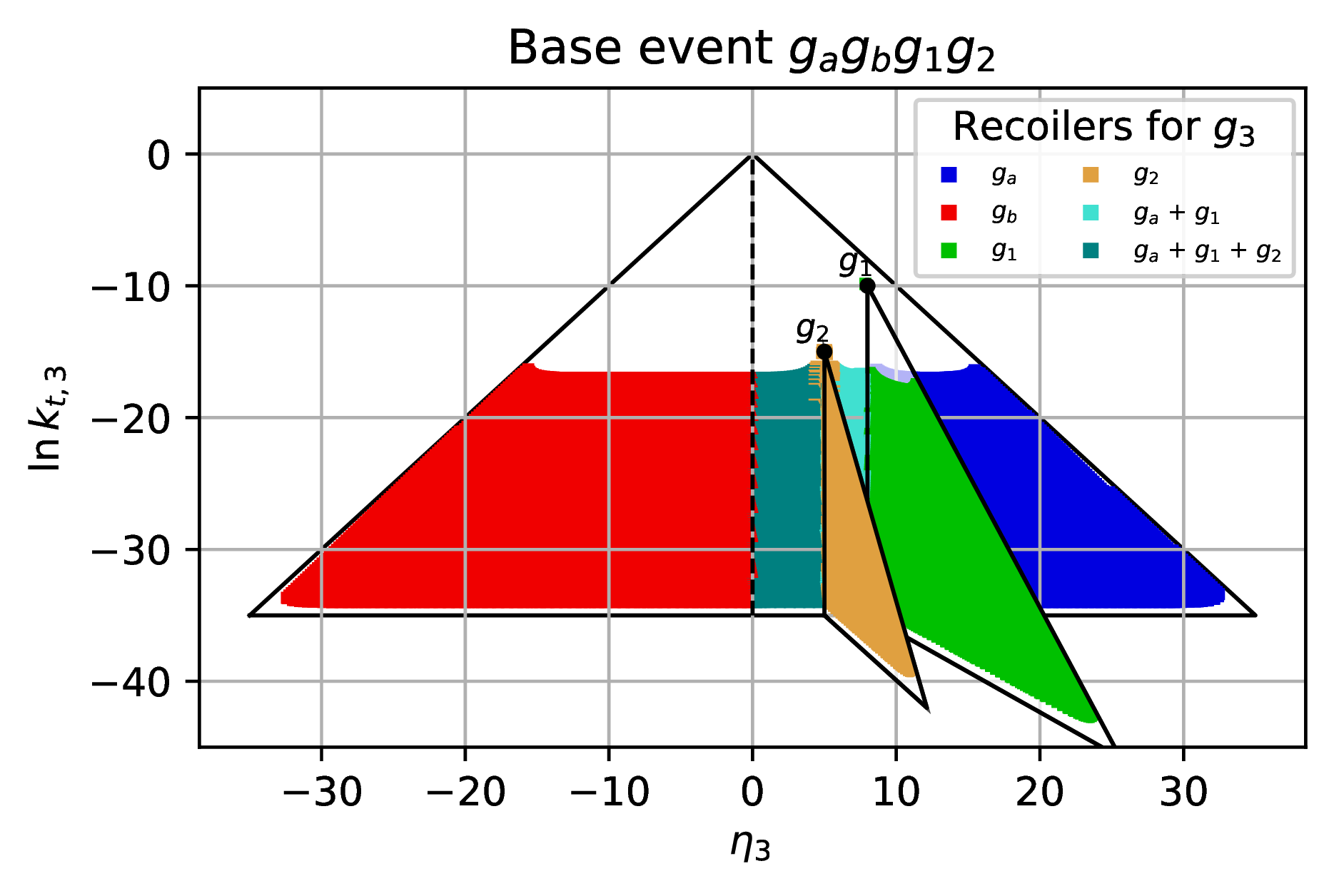}
    \includegraphics[width=.49\textwidth]{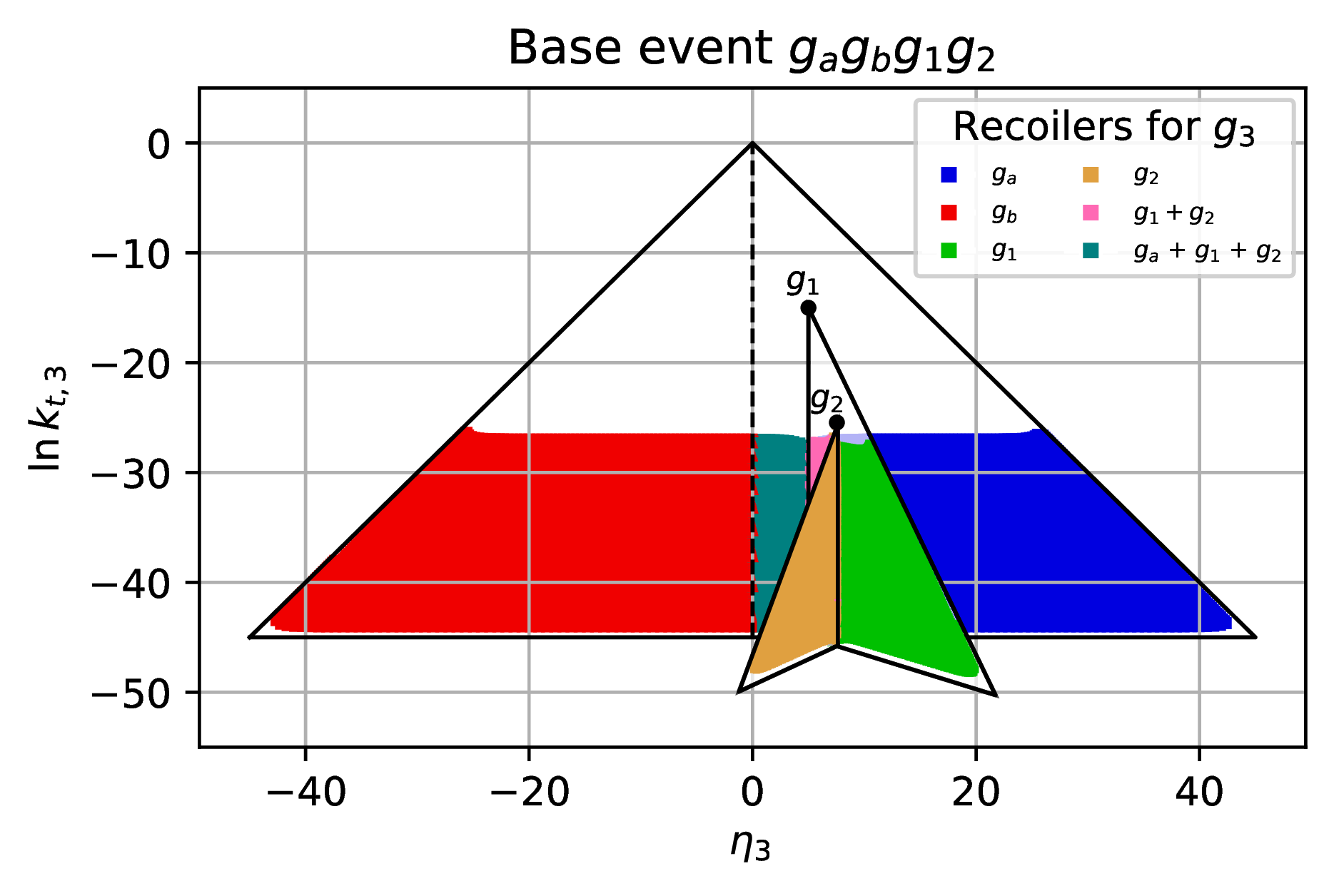}
    \caption{Fixed-order algorithmic tests of the implementation of the jet-recoil shower as in Fig.~\ref{fig:set-recoilers}, for $H \to g_ag_b$ with additional gluon emissions.}
    \label{fig:set-recoilers-H2gg}
\end{figure}

\begin{figure}[ht!]
    \centering
    \includegraphics[page=1,width=.49\textwidth]{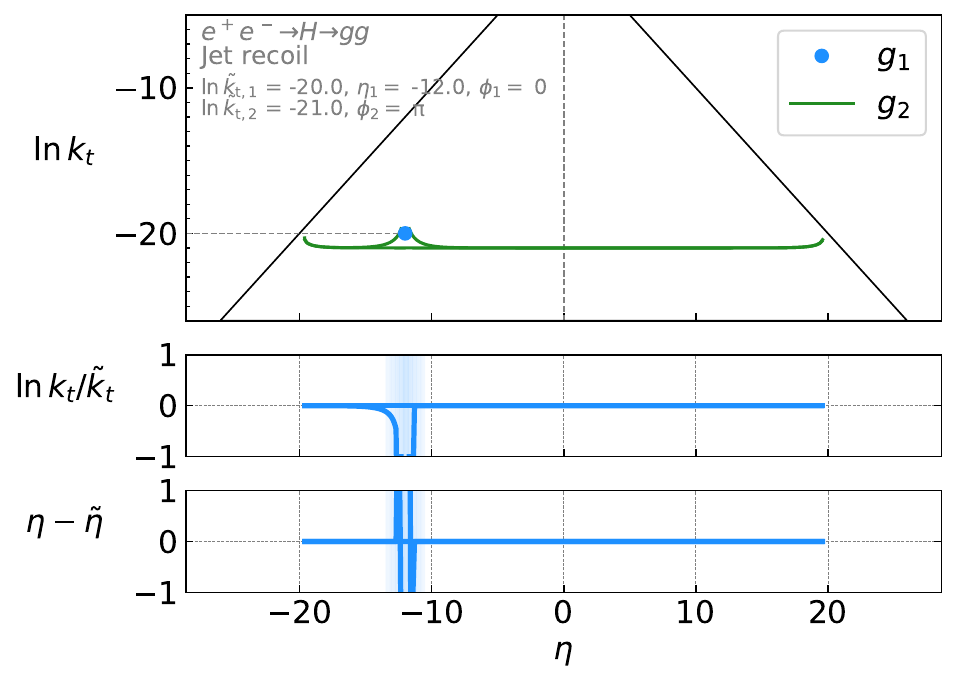}%
    \includegraphics[page=2,width=.49\textwidth]{figures/contours-H2gg.pdf}\\
    \includegraphics[page=3,width=.49\textwidth]{figures/contours-H2gg.pdf}%
    \caption{Fixed-order tests of the correct recoil assignment, as in Figs.~\ref{fig:fo-tests-0},~\ref{fig:fo-tests-1} for the jet-recoil scheme applied to $H\to gg$.}
    \label{fig:fo-tests-h2gg}
\end{figure}

\begin{figure}[ht!]
    \centering
    \includegraphics[height=5.6cm,page=4, clip]{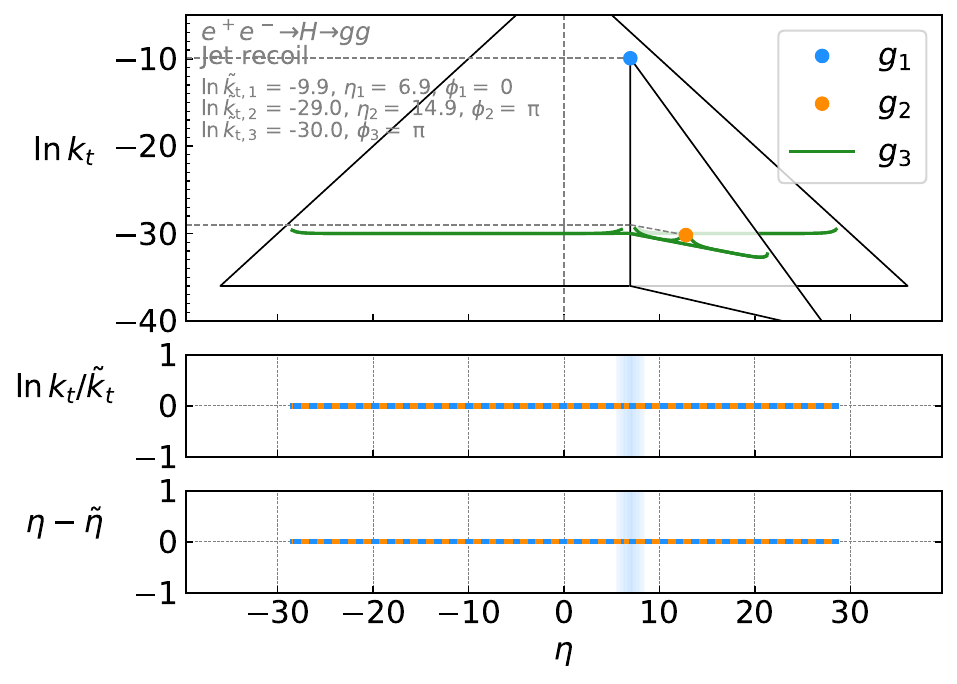}%
    \includegraphics[height=5.6cm,page=5, clip]{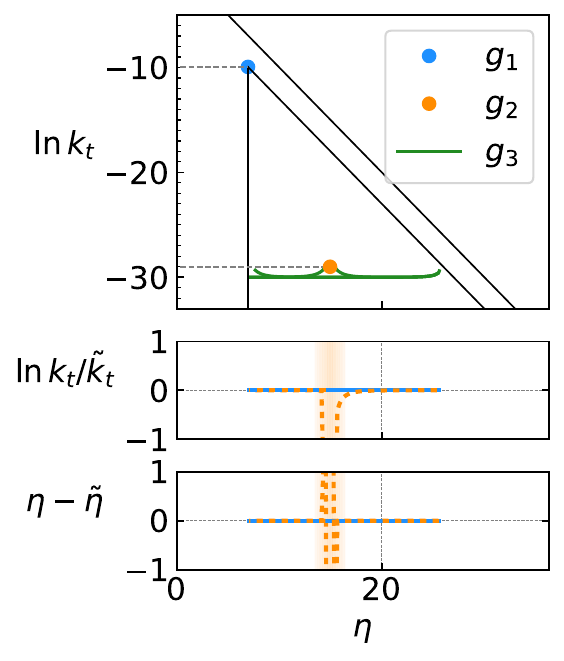}
    \caption{Same as Fig.~\ref{fig:fo-tests-sec} for $H\to gg$.}
    \label{fig:fo-tests-sec-h2gg}
\end{figure}

\begin{figure}[ht!]
    \centering
    \includegraphics[width=0.7\linewidth, clip]{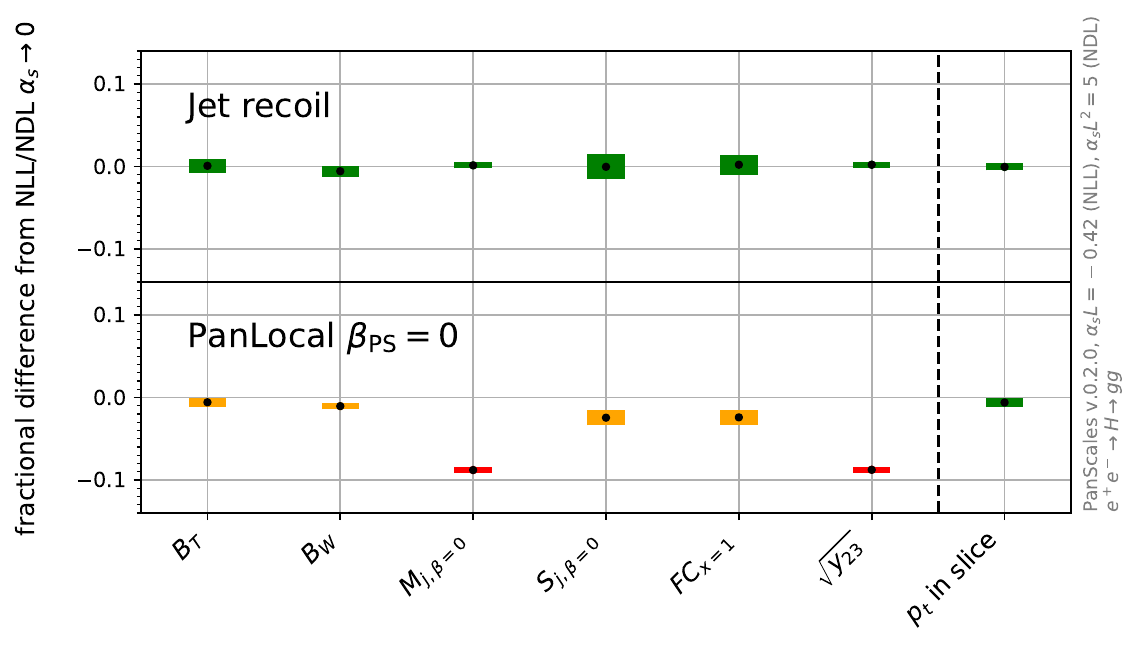}
    \caption{Same as Fig.~\ref{fig:NLL_results} for $H\to gg$, for all observables with $\betaobs=0$ and the transverse momentum in a rapidity slice.}
    \label{fig:NLL_results_H2gg}
\end{figure}

Tests of the implementation similar to those of sections~\ref{sec:fo-tests-alg},~\ref{sec:fo-tests-NLL} and~\ref{sec:nll-tests} are presented below in the case of $H\to gg$. Algorithmic tests are shown in Fig.~\ref{fig:set-recoilers-H2gg} (where the convention is that $g_a$ ($g_b$) has a positive (negative) $z$-component), and tests of the recoil for configurations at fixed order are shown in Figs.~\ref{fig:fo-tests-h2gg},~\ref{fig:fo-tests-sec-h2gg}.
Finally, because the full series of all-order logarithmic tests is computationally somewhat expensive, we run a subset of those for $H\to gg$ (namely for those observables where a mistake would lead to a visible NLL discrepancy, $\beta_{\rm obs} = 0$, and for a non-global observable). These results are presented in Fig.~\ref{fig:NLL_results_H2gg}. In all cases we see the behaviour expected of a NLL-accurate shower evolution.


\FloatBarrier
\bibliographystyle{JHEP}
\bibliography{MC}

\providecommand{\href}[2]{#2}\begingroup\raggedright\begin{thebibliography}{10}

\bibitem{Hoche:2017kst}
S.~H\"oche, D.~Reichelt and F.~Siegert, \emph{{Momentum conservation and
  unitarity in parton showers and NLL resummation}},
  \href{http://dx.doi.org/10.1007/JHEP01(2018)118}{\emph{JHEP} {\bf 01} (2018)
  118}, [\href{http://arxiv.org/abs/1711.03497}{{\tt 1711.03497}}].

\bibitem{Catani:1996vz}
S.~Catani and M.~H. Seymour, \emph{{A General algorithm for calculating jet
  cross-sections in NLO QCD}},
  \href{http://dx.doi.org/10.1016/S0550-3213(96)00589-5,
  10.1016/S0550-3213(98)81022-5}{\emph{Nucl. Phys.} {\bf B485} (1997)
  291--419}, [\href{http://arxiv.org/abs/hep-ph/9605323}{{\tt
  hep-ph/9605323}}].

\bibitem{Phaf:2001gc}
L.~Phaf and S.~Weinzierl, \emph{{Dipole formalism with heavy fermions}},
  \href{http://dx.doi.org/10.1088/1126-6708/2001/04/006}{\emph{JHEP} {\bf 04}
  (2001) 006}, [\href{http://arxiv.org/abs/hep-ph/0102207}{{\tt
  hep-ph/0102207}}].

\bibitem{Kosower:2003bh}
D.~A. Kosower, \emph{{Antenna factorization in strongly ordered limits}},
  \href{http://dx.doi.org/10.1103/PhysRevD.71.045016}{\emph{Phys. Rev. D} {\bf
  71} (2005) 045016}, [\href{http://arxiv.org/abs/hep-ph/0311272}{{\tt
  hep-ph/0311272}}].

\bibitem{Gehrmann-DeRidder:2009lyc}
A.~Gehrmann-De~Ridder and M.~Ritzmann, \emph{{NLO Antenna Subtraction with
  Massive Fermions}},
  \href{http://dx.doi.org/10.1088/1126-6708/2009/07/041}{\emph{JHEP} {\bf 07}
  (2009) 041}, [\href{http://arxiv.org/abs/0904.3297}{{\tt 0904.3297}}].

\bibitem{Gustafson:1987rq}
G.~Gustafson and U.~Pettersson, \emph{{Dipole Formulation of QCD Cascades}},
  \href{http://dx.doi.org/10.1016/0550-3213(88)90441-5}{\emph{Nucl. Phys.} {\bf
  B306} (1988) 746--758}.

\bibitem{Lonnblad:1992tz}
L.~L{\"o}nnblad, \emph{{ARIADNE version 4: A Program for simulation of QCD
  cascades implementing the color dipole model}},
  \href{http://dx.doi.org/10.1016/0010-4655(92)90068-A}{\emph{Comput. Phys.
  Commun.} {\bf 71} (1992) 15--31}.

\bibitem{Sjostrand:2004ef}
T.~Sj{\"o}strand and P.~Z. Skands, \emph{{Transverse-momentum-ordered showers
  and interleaved multiple interactions}},
  \href{http://dx.doi.org/10.1140/epjc/s2004-02084-y}{\emph{Eur. Phys. J.} {\bf
  C39} (2005) 129--154}, [\href{http://arxiv.org/abs/hep-ph/0408302}{{\tt
  hep-ph/0408302}}].

\bibitem{Bierlich:2022pfr}
C.~Bierlich et~al., \emph{{A comprehensive guide to the physics and usage of
  PYTHIA 8.3}},
  \href{http://dx.doi.org/10.21468/SciPostPhysCodeb.8}{\emph{SciPost Phys.
  Codeb.} {\bf 2022} (2022) 8}, [\href{http://arxiv.org/abs/2203.11601}{{\tt
  2203.11601}}].

\bibitem{Fischer:2016vfv}
N.~Fischer, S.~Prestel, M.~Ritzmann and P.~Skands, \emph{{Vincia for Hadron
  Colliders}},
  \href{http://dx.doi.org/10.1140/epjc/s10052-016-4429-6}{\emph{Eur. Phys. J.}
  {\bf C76} (2016) 589}, [\href{http://arxiv.org/abs/1605.06142}{{\tt
  1605.06142}}].

\bibitem{Brooks:2020upa}
H.~Brooks, C.~T. Preuss and P.~Skands, \emph{{Sector Showers for Hadron
  Collisions}}, \href{http://dx.doi.org/10.1007/JHEP07(2020)032}{\emph{JHEP}
  {\bf 07} (2020) 032}, [\href{http://arxiv.org/abs/2003.00702}{{\tt
  2003.00702}}].

\bibitem{Platzer:2011bc}
S.~Pl{\"a}tzer and S.~Gieseke, \emph{{Dipole Showers and Automated NLO Matching
  in Herwig++}},
  \href{http://dx.doi.org/10.1140/epjc/s10052-012-2187-7}{\emph{Eur. Phys. J.
  C} {\bf 72} (2012) 2187}, [\href{http://arxiv.org/abs/1109.6256}{{\tt
  1109.6256}}].

\bibitem{Bewick:2023tfi}
G.~Bewick et~al., \emph{{Herwig 7.3 release note}},
  \href{http://dx.doi.org/10.1140/epjc/s10052-024-13211-9}{\emph{Eur. Phys. J.
  C} {\bf 84} (2024) 1053}, [\href{http://arxiv.org/abs/2312.05175}{{\tt
  2312.05175}}].

\bibitem{Schumann:2007mg}
S.~Schumann and F.~Krauss, \emph{{A Parton shower algorithm based on
  Catani-Seymour dipole factorisation}},
  \href{http://dx.doi.org/10.1088/1126-6708/2008/03/038}{\emph{JHEP} {\bf 03}
  (2008) 038}, [\href{http://arxiv.org/abs/0709.1027}{{\tt 0709.1027}}].

\bibitem{Sherpa:2024mfk}
{\scshape Sherpa} collaboration, E.~Bothmann et~al., \emph{{Event generation
  with Sherpa 3}}, \href{http://dx.doi.org/10.1007/JHEP12(2024)156}{\emph{JHEP}
  {\bf 12} (2024) 156}, [\href{http://arxiv.org/abs/2410.22148}{{\tt
  2410.22148}}].

\bibitem{Hoche:2015sya}
S.~Hoeche and S.~Prestel, \emph{{The midpoint between dipole and parton
  showers}}, \href{http://dx.doi.org/10.1140/epjc/s10052-015-3684-2}{\emph{Eur.
  Phys. J.} {\bf C75} (2015) 461}, [\href{http://arxiv.org/abs/1506.05057}{{\tt
  1506.05057}}].

\bibitem{Dasgupta:2020fwr}
M.~Dasgupta, F.~A. Dreyer, K.~Hamilton, P.~F. Monni, G.~P. Salam and G.~Soyez,
  \emph{{Parton showers beyond leading logarithmic accuracy}},
  \href{http://dx.doi.org/10.1103/PhysRevLett.125.052002}{\emph{Phys. Rev.
  Lett.} {\bf 125} (2020) 052002}, [\href{http://arxiv.org/abs/2002.11114}{{\tt
  2002.11114}}].

\bibitem{Nagy:2006kb}
Z.~Nagy and D.~E. Soper, \emph{{A New parton shower algorithm: Shower
  evolution, matching at leading and next-to-leading order level}},  in
  \emph{{Ringberg Workshop on New Trends in HERA Physics 2005}}, pp.~101--123,
  1, 2006.
\newblock \href{http://arxiv.org/abs/hep-ph/0601021}{{\tt hep-ph/0601021}}.
\newblock \href{http://dx.doi.org/10.1142/9789812773524_0010}{DOI}.

\bibitem{Dinsdale:2007mf}
M.~Dinsdale, M.~Ternick and S.~Weinzierl, \emph{{Parton showers from the dipole
  formalism}}, \href{http://dx.doi.org/10.1103/PhysRevD.76.094003}{\emph{Phys.
  Rev. D} {\bf 76} (2007) 094003}, [\href{http://arxiv.org/abs/0709.1026}{{\tt
  0709.1026}}].

\bibitem{Winter:2007ye}
J.-C. Winter and F.~Krauss, \emph{{Initial-state showering based on colour
  dipoles connected to incoming parton lines}},
  \href{http://dx.doi.org/10.1088/1126-6708/2008/07/040}{\emph{JHEP} {\bf 07}
  (2008) 040}, [\href{http://arxiv.org/abs/0712.3913}{{\tt 0712.3913}}].

\bibitem{Marchesini:1983bm}
G.~Marchesini and B.~R. Webber, \emph{{Simulation of QCD Jets Including Soft
  Gluon Interference}},
  \href{http://dx.doi.org/10.1016/0550-3213(84)90463-2}{\emph{Nucl. Phys.} {\bf
  B238} (1984) 1--29}.

\bibitem{Gieseke:2003rz}
S.~Gieseke, P.~Stephens and B.~Webber, \emph{{New formalism for QCD parton
  showers}}, \href{http://dx.doi.org/10.1088/1126-6708/2003/12/045}{\emph{JHEP}
  {\bf 12} (2003) 045}, [\href{http://arxiv.org/abs/hep-ph/0310083}{{\tt
  hep-ph/0310083}}].

\bibitem{Bewick:2019rbu}
G.~Bewick, S.~Ferrario~Ravasio, P.~Richardson and M.~H. Seymour,
  \emph{{Logarithmic accuracy of angular-ordered parton showers}},
  \href{http://dx.doi.org/10.1007/JHEP04(2020)019}{\emph{JHEP} {\bf 04} (2020)
  019}, [\href{http://arxiv.org/abs/1904.11866}{{\tt 1904.11866}}].

\bibitem{Forshaw:2020wrq}
J.~R. Forshaw, J.~Holguin and S.~Pl\"atzer, \emph{{Building a consistent parton
  shower}}, \href{http://dx.doi.org/10.1007/JHEP09(2020)014}{\emph{JHEP} {\bf
  09} (2020) 014}, [\href{http://arxiv.org/abs/2003.06400}{{\tt 2003.06400}}].

\bibitem{Nagy:2014mqa}
Z.~Nagy and D.~E. Soper, \emph{{A parton shower based on factorization of the
  quantum density matrix}},
  \href{http://dx.doi.org/10.1007/JHEP06(2014)097}{\emph{JHEP} {\bf 06} (2014)
  097}, [\href{http://arxiv.org/abs/1401.6364}{{\tt 1401.6364}}].

\bibitem{Nagy:2020rmk}
Z.~Nagy and D.~E. Soper, \emph{{Summations of large logarithms by parton
  showers}}, \href{http://dx.doi.org/10.1103/PhysRevD.104.054049}{\emph{Phys.
  Rev. D} {\bf 104} (2021) 054049},
  [\href{http://arxiv.org/abs/2011.04773}{{\tt 2011.04773}}].

\bibitem{FerrarioRavasio:2023kyg}
S.~Ferrario~Ravasio, K.~Hamilton, A.~Karlberg, G.~P. Salam, L.~Scyboz and
  G.~Soyez, \emph{{Parton Showering with Higher Logarithmic Accuracy for Soft
  Emissions}},
  \href{http://dx.doi.org/10.1103/PhysRevLett.131.161906}{\emph{Phys. Rev.
  Lett.} {\bf 131} (2023) 161906}, [\href{http://arxiv.org/abs/2307.11142}{{\tt
  2307.11142}}].

\bibitem{Herren:2022jej}
F.~Herren, S.~H\"oche, F.~Krauss, D.~Reichelt and M.~Schoenherr, \emph{{A new
  approach to color-coherent parton evolution}},
  \href{http://dx.doi.org/10.1007/JHEP10(2023)091}{\emph{JHEP} {\bf 10} (2023)
  091}, [\href{http://arxiv.org/abs/2208.06057}{{\tt 2208.06057}}].

\bibitem{Assi:2023rbu}
B.~Assi and S.~H\"oche, \emph{{New approach to QCD final-state evolution in
  processes with massive partons}},
  \href{http://dx.doi.org/10.1103/PhysRevD.109.114008}{\emph{Phys. Rev. D} {\bf
  109} (2024) 114008}, [\href{http://arxiv.org/abs/2307.00728}{{\tt
  2307.00728}}].

\bibitem{Hoche:2024dee}
S.~H{\"o}che, F.~Krauss and D.~Reichelt, \emph{{The Alaric parton shower for
  hadron colliders}},
  \href{http://dx.doi.org/10.1103/PhysRevD.111.094032}{\emph{Phys. Rev. D} {\bf
  111} (2025) 094032}, [\href{http://arxiv.org/abs/2404.14360}{{\tt
  2404.14360}}].

\bibitem{Preuss:2024vyu}
C.~T. Preuss, \emph{{A partitioned dipole-antenna shower with improved
  transverse recoil}},
  \href{http://dx.doi.org/10.1007/JHEP07(2024)161}{\emph{JHEP} {\bf 07} (2024)
  161}, [\href{http://arxiv.org/abs/2403.19452}{{\tt 2403.19452}}].

\bibitem{Dasgupta:2018nvj}
M.~Dasgupta, F.~A. Dreyer, K.~Hamilton, P.~F. Monni and G.~P. Salam,
  \emph{{Logarithmic accuracy of parton showers: a fixed-order study}},
  \href{http://dx.doi.org/10.1007/JHEP09(2018)033}{\emph{JHEP} {\bf 09} (2018)
  033}, [\href{http://arxiv.org/abs/1805.09327}{{\tt 1805.09327}}].

\bibitem{Banfi:2006gy}
A.~Banfi, G.~Corcella and M.~Dasgupta, \emph{{Angular ordering and parton
  showers for non-global QCD observables}},
  \href{http://dx.doi.org/10.1088/1126-6708/2007/03/050}{\emph{JHEP} {\bf 03}
  (2007) 050}, [\href{http://arxiv.org/abs/hep-ph/0612282}{{\tt
  hep-ph/0612282}}].

\bibitem{vanBeekveld:2022ukn}
M.~van Beekveld, S.~Ferrario~Ravasio, K.~Hamilton, G.~P. Salam, A.~Soto-Ontoso,
  G.~Soyez et~al., \emph{{PanScales showers for hadron collisions: all-order
  validation}}, \href{http://dx.doi.org/10.1007/JHEP11(2022)020}{\emph{JHEP}
  {\bf 11} (2022) 020}, [\href{http://arxiv.org/abs/2207.09467}{{\tt
  2207.09467}}].

\bibitem{vanBeekveld:2024wws}
M.~van Beekveld et~al., \emph{{New Standard for the Logarithmic Accuracy of
  Parton Showers}},
  \href{http://dx.doi.org/10.1103/PhysRevLett.134.011901}{\emph{Phys. Rev.
  Lett.} {\bf 134} (2025) 011901}, [\href{http://arxiv.org/abs/2406.02661}{{\tt
  2406.02661}}].

\bibitem{SalamSoyez:unpublished}
G.~Salam and G.~Soyez, \emph{Unpublished notes},  2022.

\bibitem{Gribov:1972ri}
V.~N. Gribov and L.~N. Lipatov, \emph{{Deep inelastic e p scattering in
  perturbation theory}}, {\emph{Sov. J. Nucl. Phys.} {\bf 15} (1972) 438--450}.

\bibitem{Altarelli:1977zs}
G.~Altarelli and G.~Parisi, \emph{{Asymptotic Freedom in Parton Language}},
  \href{http://dx.doi.org/10.1016/0550-3213(77)90384-4}{\emph{Nucl. Phys.} {\bf
  B126} (1977) 298--318}.

\bibitem{Dokshitzer:1977sg}
Y.~L. Dokshitzer, \emph{{Calculation of the Structure Functions for Deep
  Inelastic Scattering and e+ e- Annihilation by Perturbation Theory in Quantum
  Chromodynamics.}}, {\emph{Sov. Phys. JETP} {\bf 46} (1977) 641--653}.

\bibitem{Catani:1996jh}
S.~Catani and M.~H. Seymour, \emph{{The Dipole formalism for the calculation of
  QCD jet cross-sections at next-to-leading order}},
  \href{http://dx.doi.org/10.1016/0370-2693(96)00425-X}{\emph{Phys. Lett. B}
  {\bf 378} (1996) 287--301}, [\href{http://arxiv.org/abs/hep-ph/9602277}{{\tt
  hep-ph/9602277}}].

\bibitem{Kosower:1997zr}
D.~A. Kosower, \emph{{Antenna factorization of gauge theory amplitudes}},
  \href{http://dx.doi.org/10.1103/PhysRevD.57.5410}{\emph{Phys. Rev. D} {\bf
  57} (1998) 5410--5416}, [\href{http://arxiv.org/abs/hep-ph/9710213}{{\tt
  hep-ph/9710213}}].

\bibitem{Gehrmann-DeRidder:2005btv}
A.~Gehrmann-De~Ridder, T.~Gehrmann and E.~W.~N. Glover, \emph{{Antenna
  subtraction at NNLO}},
  \href{http://dx.doi.org/10.1088/1126-6708/2005/09/056}{\emph{JHEP} {\bf 09}
  (2005) 056}, [\href{http://arxiv.org/abs/hep-ph/0505111}{{\tt
  hep-ph/0505111}}].

\bibitem{Banfi:2001bz}
A.~Banfi, G.~P. Salam and G.~Zanderighi, \emph{{Semi-numerical resummation of
  event shapes}},
  \href{http://dx.doi.org/10.1088/1126-6708/2002/01/018}{\emph{JHEP} {\bf 01}
  (2002) 018}, [\href{http://arxiv.org/abs/hep-ph/0112156}{{\tt
  hep-ph/0112156}}].

\bibitem{Banfi:2004yd}
A.~Banfi, G.~P. Salam and G.~Zanderighi, \emph{{Principles of general
  final-state resummation and automated implementation}},
  \href{http://dx.doi.org/10.1088/1126-6708/2005/03/073}{\emph{JHEP} {\bf 03}
  (2005) 073}, [\href{http://arxiv.org/abs/hep-ph/0407286}{{\tt
  hep-ph/0407286}}].

\bibitem{Catani:1992ua}
S.~Catani, L.~Trentadue, G.~Turnock and B.~R. Webber, \emph{{Resummation of
  large logarithms in e+ e- event shape distributions}},
  \href{http://dx.doi.org/10.1016/0550-3213(93)90271-P}{\emph{Nucl. Phys.} {\bf
  B407} (1993) 3--42}.

\bibitem{Andersson:1991he}
B.~Andersson, G.~Gustafson and C.~Sjogren, \emph{{Comparison of the dipole
  cascade model versus $O(\alpha_s^2)$ matrix elements and color interference
  in $e^+e^-$ annihilation}},
  \href{http://dx.doi.org/10.1016/0550-3213(92)90250-F}{\emph{Nucl. Phys.} {\bf
  B380} (1992) 391--407}.

\bibitem{vanBeekveld:2022zhl}
M.~van Beekveld, S.~Ferrario~Ravasio, G.~P. Salam, A.~Soto-Ontoso, G.~Soyez and
  R.~Verheyen, \emph{{PanScales parton showers for hadron collisions:
  formulation and fixed-order studies}},
  \href{http://dx.doi.org/10.1007/JHEP11(2022)019}{\emph{JHEP} {\bf 11} (2022)
  019}, [\href{http://arxiv.org/abs/2205.02237}{{\tt 2205.02237}}].

\bibitem{vanBeekveld:2023chs}
M.~van Beekveld and S.~Ferrario~Ravasio, \emph{{Next-to-leading-logarithmic
  PanScales showers for Deep Inelastic Scattering and Vector Boson Fusion}},
  \href{http://dx.doi.org/10.1007/JHEP02(2024)001}{\emph{JHEP} {\bf 02} (2024)
  001}, [\href{http://arxiv.org/abs/2305.08645}{{\tt 2305.08645}}].

\bibitem{Andersson:1988gp}
B.~Andersson, G.~Gustafson, L.~L{\"o}nnblad and U.~Pettersson, \emph{{Coherence
  Effects in Deep Inelastic Scattering}},
  \href{http://dx.doi.org/10.1007/BF01550942}{\emph{Z. Phys.} {\bf C43} (1989)
  625}.

\bibitem{Dreyer:2018nbf}
F.~A. Dreyer, G.~P. Salam and G.~Soyez, \emph{{The Lund Jet Plane}},
  \href{http://dx.doi.org/10.1007/JHEP12(2018)064}{\emph{JHEP} {\bf 12} (2018)
  064}, [\href{http://arxiv.org/abs/1807.04758}{{\tt 1807.04758}}].

\bibitem{Dokshitzer:1997in}
Y.~L. Dokshitzer, G.~D. Leder, S.~Moretti and B.~R. Webber, \emph{{Better jet
  clustering algorithms}},
  \href{http://dx.doi.org/10.1088/1126-6708/1997/08/001}{\emph{JHEP} {\bf 08}
  (1997) 001}, [\href{http://arxiv.org/abs/hep-ph/9707323}{{\tt
  hep-ph/9707323}}].

\bibitem{Cacciari:2011ma}
M.~Cacciari, G.~P. Salam and G.~Soyez, \emph{{FastJet User Manual}},
  \href{http://dx.doi.org/10.1140/epjc/s10052-012-1896-2}{\emph{Eur. Phys. J.}
  {\bf C72} (2012) 1896}, [\href{http://arxiv.org/abs/1111.6097}{{\tt
  1111.6097}}].

\bibitem{Mueller:1981ex}
A.~H. Mueller, \emph{{On the Multiplicity of Hadrons in QCD Jets}},
  \href{http://dx.doi.org/10.1016/0370-2693(81)90581-5}{\emph{Phys. Lett. B}
  {\bf 104} (1981) 161--164}.

\bibitem{Mueller:1983js}
A.~H. Mueller, \emph{{Multiplicity and Hadron Distributions in {QCD} Jets. 2. A
  General Procedure for All Nonleading Terms}},
  \href{http://dx.doi.org/10.1016/0550-3213(83)90329-2}{\emph{Nucl. Phys. B}
  {\bf 228} (1983) 351--364}.

\bibitem{Ermolaev:1981cm}
B.~I. Ermolaev and V.~S. Fadin, \emph{{Log - Log Asymptotic Form of Exclusive
  Cross-Sections in Quantum Chromodynamics}}, {\emph{JETP Lett.} {\bf 33}
  (1981) 269--272}.

\bibitem{Bassetto:1982ma}
A.~Bassetto, M.~Ciafaloni, G.~Marchesini and A.~H. Mueller, \emph{{Jet
  Multiplicity and Soft Gluon Factorization}},
  \href{http://dx.doi.org/10.1016/0550-3213(82)90161-4}{\emph{Nucl. Phys. B}
  {\bf 207} (1982) 189--204}.

\bibitem{Dokshitzer:1982xr}
Y.~L. Dokshitzer, V.~S. Fadin and V.~A. Khoze, \emph{{Double Logs of
  Perturbative QCD for Parton Jets and Soft Hadron Spectra}},
  \href{http://dx.doi.org/10.1007/BF01614423}{\emph{Z. Phys. C} {\bf 15} (1982)
  325}.

\bibitem{Dokshitzer:1982ia}
Y.~L. Dokshitzer, V.~S. Fadin and V.~A. Khoze, \emph{{On the Sensitivity of the
  Inclusive Distributions in Parton Jets to the Coherence Effects in QCD Gluon
  Cascades}}, \href{http://dx.doi.org/10.1007/BF01571703}{\emph{Z. Phys. C}
  {\bf 18} (1983) 37}.

\bibitem{Hamilton:2020rcu}
K.~Hamilton, R.~Medves, G.~P. Salam, L.~Scyboz and G.~Soyez, \emph{{Colour and
  logarithmic accuracy in final-state parton showers}},
  \href{http://dx.doi.org/10.1007/JHEP03(2021)041}{\emph{JHEP} {\bf 03} (2021)
  041}, [\href{http://arxiv.org/abs/2011.10054}{{\tt 2011.10054}}].

\bibitem{vanBeekveld:2023ivn}
M.~van Beekveld et~al., \emph{{Introduction to the PanScales framework, version
  0.1}}, \href{http://dx.doi.org/10.21468/SciPostPhysCodeb.31}{\emph{SciPost
  Phys. Codeb.} {\bf 2024} (2024) 31},
  [\href{http://arxiv.org/abs/2312.13275}{{\tt 2312.13275}}].

\bibitem{Catani:1990rr}
S.~Catani, B.~R. Webber and G.~Marchesini, \emph{{QCD coherent branching and
  semiinclusive processes at large x}},
  \href{http://dx.doi.org/10.1016/0550-3213(91)90390-J}{\emph{Nucl. Phys.} {\bf
  B349} (1991) 635--654}.

\bibitem{Dasgupta:2001sh}
M.~Dasgupta and G.~Salam, \emph{{Resummation of nonglobal QCD observables}},
  \href{http://dx.doi.org/10.1016/S0370-2693(01)00725-0}{\emph{Phys. Lett. B}
  {\bf 512} (2001) 323--330}, [\href{http://arxiv.org/abs/hep-ph/0104277}{{\tt
  hep-ph/0104277}}].

\bibitem{Dasgupta:2002bw}
M.~Dasgupta and G.~P. Salam, \emph{{Accounting for coherence in interjet E(t)
  flow: A Case study}},
  \href{http://dx.doi.org/10.1088/1126-6708/2002/03/017}{\emph{JHEP} {\bf 03}
  (2002) 017}, [\href{http://arxiv.org/abs/hep-ph/0203009}{{\tt
  hep-ph/0203009}}].

\bibitem{Catani:1991pm}
S.~Catani, Y.~L. Dokshitzer, F.~Fiorani and B.~R. Webber, \emph{{Average number
  of jets in $e^+ e^-$ annihilation}},
  \href{http://dx.doi.org/10.1016/0550-3213(92)90296-N}{\emph{Nucl. Phys.} {\bf
  B377} (1992) 445--460}.

\bibitem{Medves:2022ccw}
R.~Medves, A.~Soto-Ontoso and G.~Soyez, \emph{{Lund and Cambridge
  multiplicities for precision physics}},
  \href{http://dx.doi.org/10.1007/JHEP10(2022)156}{\emph{JHEP} {\bf 10} (2022)
  156}, [\href{http://arxiv.org/abs/2205.02861}{{\tt 2205.02861}}].

\bibitem{Catani:1992jc}
S.~Catani, G.~Turnock and B.~R. Webber, \emph{{Jet broadening measures in
  $e^{+} e^{-}$ annihilation}},
  \href{http://dx.doi.org/10.1016/0370-2693(92)91565-Q}{\emph{Phys. Lett.} {\bf
  B295} (1992) 269--276}.

\bibitem{Brandt:1964sa}
S.~Brandt, C.~Peyrou, R.~Sosnowski and A.~Wroblewski, \emph{{The Principal axis
  of jets. An Attempt to analyze high-energy collisions as two-body
  processes}},
  \href{http://dx.doi.org/10.1016/0031-9163(64)91176-X}{\emph{Phys. Lett.} {\bf
  12} (1964) 57--61}.

\bibitem{Farhi:1977sg}
E.~Farhi, \emph{{A QCD Test for Jets}},
  \href{http://dx.doi.org/10.1103/PhysRevLett.39.1587}{\emph{Phys. Rev. Lett.}
  {\bf 39} (1977) 1587--1588}.

\bibitem{Catani:1991hj}
S.~Catani, Y.~L. Dokshitzer, M.~Olsson, G.~Turnock and B.~R. Webber, \emph{{New
  clustering algorithm for multi - jet cross-sections in $e^+ e^-$
  annihilation}},
  \href{http://dx.doi.org/10.1016/0370-2693(91)90196-W}{\emph{Phys. Lett.} {\bf
  B269} (1991) 432--438}.

\bibitem{Collins:1987cp}
J.~C. Collins, \emph{{Spin Correlations in Monte Carlo Event Generators}},
  \href{http://dx.doi.org/10.1016/0550-3213(88)90654-2}{\emph{Nucl. Phys.} {\bf
  B304} (1988) 794--804}.

\bibitem{Knowles:1988hu}
I.~G. Knowles, \emph{{A Linear Algorithm for Calculating Spin Correlations in
  Hadronic Collisions}},
  \href{http://dx.doi.org/10.1016/0010-4655(90)90063-7}{\emph{Comput. Phys.
  Commun.} {\bf 58} (1990) 271--284}.

\bibitem{Hoche:2025anb}
S.~H{\"o}che, M.~Hoppe and D.~Reichelt, \emph{{A simple algorithm for polarized
  parton evolution}},  \href{http://arxiv.org/abs/2508.19018}{{\tt
  2508.19018}}.

\bibitem{Richardson:2001df}
P.~Richardson, \emph{{Spin correlations in Monte Carlo simulations}},
  \href{http://dx.doi.org/10.1088/1126-6708/2001/11/029}{\emph{JHEP} {\bf 11}
  (2001) 029}, [\href{http://arxiv.org/abs/hep-ph/0110108}{{\tt
  hep-ph/0110108}}].

\bibitem{Richardson:2018pvo}
P.~Richardson and S.~Webster, \emph{{Spin Correlations in Parton Shower
  Simulations}},
  \href{http://dx.doi.org/10.1140/epjc/s10052-019-7429-5}{\emph{Eur. Phys. J.}
  {\bf C80} (2020) 83}, [\href{http://arxiv.org/abs/1807.01955}{{\tt
  1807.01955}}].

\bibitem{Karlberg:2021kwr}
A.~Karlberg, G.~P. Salam, L.~Scyboz and R.~Verheyen, \emph{{Spin correlations
  in final-state parton showers and jet observables}},
  \href{http://dx.doi.org/10.1140/epjc/s10052-021-09378-0}{\emph{Eur. Phys. J.
  C} {\bf 81} (2021) 681}, [\href{http://arxiv.org/abs/2103.16526}{{\tt
  2103.16526}}].

\bibitem{Hamilton:2021dyz}
K.~Hamilton, A.~Karlberg, G.~P. Salam, L.~Scyboz and R.~Verheyen, \emph{{Soft
  spin correlations in final-state parton showers}},
  \href{http://dx.doi.org/10.1007/JHEP03(2022)193}{\emph{JHEP} {\bf 03} (2022)
  193}, [\href{http://arxiv.org/abs/2111.01161}{{\tt 2111.01161}}].

\bibitem{WiltonThesis}
W.~Deany, \emph{Antenna showers with jet recoils},  Honours Thesis, School of
  Physics and Astronomy, Monash University, 2023.

\end{thebibliography}\endgroup

\end{document}